 \documentclass[a4paper]{article}
 \usepackage{etoolbox}
 \usepackage{abstract}
 \newbool{PREPRINT}
 \booltrue{PREPRINT}
\usepackage{makecell}
\usepackage[colorlinks,bookmarksopen,bookmarksnumbered,citecolor=red,urlcolor=red]{hyperref}
\usepackage{stmaryrd}
\usepackage[usenames]{color}
\usepackage{comment}
\usepackage{amsmath,amssymb}
\usepackage{graphicx}
\usepackage{algorithm}
\usepackage{algorithmicx}
\usepackage{algcompatible}
\usepackage{caption}
\usepackage{listings}
\usepackage{mathtools}
\usepackage{multicol}
\usepackage{placeins}
\usepackage{multirow}
\usepackage{subcaption}
\usepackage{enumitem}
\setlist[description]{itemsep=-0.5ex}
\newcommand{\figdir}{./}

\renewcommand{\vec}[1]{\boldsymbol{#1}}

\newif\ifpreprint
\ifbool{PREPRINT}{ 
\preprinttrue
\usepackage[cm]{fullpage}
\usepackage{amsthm}
\usepackage{multicol}
\setlength\columnsep{20pt}
\usepackage[affil-it]{authblk}

}{ %
\usepackage{fullpage}
\newdefinition{defn}{Definition}
\newdefinition{expl}{Example}
} 
\newcommand{\lr}{\infty}
\newcommand{\sr}{\square}

\newcommand{\partidxi}{i}
\newcommand{\partidxj}{j}

\newcommand{\cellidxi}{\alpha}
\newcommand{\cellidxj}{\beta}

\newcommand{\stateidxi}{\mathfrak{a}}
\newcommand{\stateidxj}{\mathfrak{b}}
\newcommand{\stateidxk}{\mathfrak{c}}

\newcommand{\Nstates}{\mathfrak{N}}

\newcommand{\ooverline}[1]{\overline{\overline{#1}}}

\lstdefinelanguage[ppmd]{python}[]{python}{%
  emph={ParticleLoop,ParticleDat,PositionDat,ScalarArray,GlobalArray,Kernel,PairLoop,Constant,State,Data,IntegratorRange}
}
\definecolor{DarkBlue}{rgb}{0.00,0.00,0.55}
\definecolor{DarkRed}{rgb}{0.55,0.00,0.00}
\definecolor{DarkGreen}{rgb}{0.00,0.55,0.00}
\definecolor{Gray}{rgb}{0.95,0.95,0.95}
\definecolor{Purple}{rgb}{0.5,0.0,0.5}
\definecolor{Bittersweet}{rgb}{1.0,0.44,0.37}
\newcommand{\secapp}{\ifbool{PREPRINT}{Appendix~}{}}
\newcommand{\pparagraph}[1]{\ifbool{PREPRINT}{\paragraph{#1.}}{\paragraph{#1}}}

\title{Fast electrostatic solvers for kinetic Monte Carlo simulations}
\ifbool{PREPRINT}{ 
  \author[a,c]{William~Robert~Saunders}
  \author[b]{James~Grant}
  \author[a,*]{Eike~Hermann~M\"{u}ller}
  \author[c]{Ian~Thompson}
  \affil[ ]{University of Bath, Bath BA2 7AY, Bath, United Kingdom}
  \affil[a]{Department of Mathematical Sciences}
  \affil[b]{Computing Services}
  \affil[c]{Department of Physics}
\affil[*]{Email: \texttt{e.mueller@bath.ac.uk}}
}{ %
\author[math,phys]{William~Robert~Saunders}
\ead{w.r.saunders@bath.ac.uk}
\author[comp]{James~Grant}
\ead{r.j.grant@bath.ac.uk}
\author[math]{Eike~Hermann~M\"{u}ller\corref{cor1}\fnref{fn1}}
\ead{e.mueller@bath.ac.uk}
\author[phys]{Ian~Thompson}
\ead{i.r.thompson@bath.ac.uk}
\fntext[fn1]{email: \texttt{e.mueller@bath.ac.uk}}
\cortext[cor1]{Corresponding author}
\address{University of Bath, Bath BA2 7AY, Bath, United Kingdom}
\address[math]{Department of Mathematical Sciences}
\address[comp]{Computing Services}
\address[phys]{Department of Physics}
} 

\begin{document}
\ifbool{PREPRINT}{ 
  \maketitle
  \begin{abstract}
}{%
\begin{abstract}
} 
Kinetic Monte Carlo (KMC) is an important computational tool in theoretical physics and chemistry. In contrast to standard Monte Carlo, KMC permits the description of time dependent dynamical processes and is not restricted to systems in equilibrium. Compared to Molecular Dynamics, it allows simulations over significantly longer timescales. Recently KMC has been applied successfully in modelling of novel energy materials such as Lithium-ion batteries and organic/perovskite solar cells. Motivated by this, we consider general solid state systems which contain free, interacting particles which can hop between localised sites in the material. The KMC transition rates for those hops depend on the change in total potential energy of the system. For charged particles this requires the frequent calculation of electrostatic interactions, which is usually the bottleneck of the simulation. To avoid this issue and obtain results in reasonable times, many studies replace the long-range potential by a phenomenological short range approximation. This, however, leads to systematic errors and unphysical results. On the other hand standard electrostatic solvers such as Ewald summation or fast Poisson solvers are highly inefficient in the KMC setup or introduce uncontrollable systematic errors at high resolution. 

In this paper we describe how the Fast Multipole Method by Greengard and Rokhlin can be adapted to overcome this issue by dramatically reducing computational costs. We exploit the fact that each update in the transition rate calculation corresponds to a single particle move and changes the configuration only by a small amount. This allows us to construct an algorithm which scales linearly in the number of charges for each KMC step, something which had not been deemed to be possible before.

We demonstrate the performance and parallel scalability of the method by implementing it in a performance portable software library, which was recently developed in our group. We describe the high-level Python interface of the code which makes it easy to adapt to specific use cases.
\ifbool{PREPRINT}{ 
  \end{abstract}

\textbf{keywords}:
\newcommand{\sep}{, }
}{
\end{abstract}
\begin{keyword}
} 
kinetic Monte Carlo\sep electrostatics\sep Fast Multipole Method\sep parallel computing\sep Domain Specific Language
\ifbool{PREPRINT}{ 
\\[1ex]
}{
\end{keyword}
\maketitle
}
\ifpreprint
\else
\fi

\section{Introduction}
The kinetic Monte Carlo (KMC) method \cite{Young1966,Bortz1975,Gillespie1976,Gillespie1977} was originally developed for the simulation of time dependent statistical processes in chemical reaction dynamics. More recently, the method has been applied in computational physics and chemistry to model processes on grain surfaces \cite{Cuppen2018}, in electrolytes \cite{Morgan2017} and organic devices \cite{Groves2016,Thompson2018}. In contrast to standard Monte Carlo methods, such as the Metropolis Hastings algorithm \cite{Metropolis1953,Hastings1970} for the simulation of Markov processes, KMC is not limited to systems in equilibrium. Instead, it allows the representation of dynamical processes in physical materials, while not being limited by the restrictive timestep constraints in Molecular Dynamics (MD) simulations which arise due to fast, but dynamically irrelevant, oscillations around semi-stable configurations.
\pparagraph{Kinetic Monte Carlo for energy materials}
An important application of the KMC method, which is currently attracting significant interest, is the simulation of transport processes in energy materials. This includes solid state electrolytes such as Lithium-ion batteries \cite{Morgan2017} and semiconductors in full-device simulations \cite{Groves2016,Thompson2018}. A particularly promising application are organic- and perovskite- based solar cells, which can achieve remarkable power efficiencies \cite{Meng2018,Peplow2018}. All those applications require the modelling of dynamic transport processes in a three dimensional volume to predict material properties such charge mobilities and the current-voltage characteristics.

In addition to making direct physical predictions, KMC simulations are also important to adjust parameters in large-scale drift-diffusion models via upscaling. Continuum models of this type are widely used in industry for full-device simulations.

The dynamics of a wide class of systems can be described by hopping processes of particles between sites of a static background matrix. To resolve physically relevant macroscopic features, such as grain boundaries, the simulation domain has to be large and simulated systems have to contain $N=10^3-10^6$ particles and hopping sites. The hopping rates (commonly referred to as ``propensities'' in the KMC literature), which serve as an input to the KMC algorithm, depend on the total potential energy of the system. This energy is given by the sum of classical interaction potentials for all particle pairs.
\pparagraph{Electrostatic interactions}
For charged particles the electrostatic contribution to the inter-particle potential is long-range. The Coulomb interactions between all particles need to be computed, resulting naively in an expensive $\mathcal{O}(N^2)$ calculation \textit{per potential hopping event}. This computational complexity can be reduced to $\mathcal{O}(N^{3/2})$ with Ewald summation \cite{Ewald1921} and $\mathcal{O}(N\log(N))$ with particle mesh methods \cite{Hockney1988,Deserno1998}. However, even if the Fast Multipole Method (FMM) \cite{Greengard1987,Greengard1988,Greengard1997} is used (in its standard form), the computational complexity \textit{per hop} is still $\mathcal{O}(N)$. Since $\mathcal{O}(N)$ potential hopping events have to be considered in each KMC step, the total cost per step is at least $\mathcal{O}(N^2)$. Because of this quadratic growth in complexity, the electrostatic calculation is typically the bottleneck of the KMC simulation. As argued in \cite{Li2018}, this appears to make large-scale KMC simulations with accurate electrostatics computationally infeasible in principle. To overcome this issue, the Coulomb potential is often replaced by an ad-hoc truncated short range interaction, see e.g. \cite{Morgan2017}. Since only the interactions with a small number of neighbours need to be calculated in this case, the cost of each potential hopping event is reduced to $\mathcal{O}(1)$, resulting in a total cost of $\mathcal{O}(N)$ per KMC step. However, this truncation introduces uncontrollable systematic errors, which limit the predictive power of the model \cite{Casalegno2010}. For example, the authors of \cite{Hermet2013} find that neglecting long range interactions when modelling protonic diffusion and conduction in doped perovskites changes the predicted diffusion coefficient by $14\%$ compared to the ``correct'' results obtained with the very expensive Ewald method. As we will discuss below, other approximation methods such as mapping the charges to a grid and solving the Poisson equation \cite{Li2017}, possibly in lower dimensions \cite{vanderHolst2011,Kordt2015}, also introduces uncontrollable systematic errors.
\pparagraph{An algorithmically optimal Fast Multipole Method}
In this paper we introduce a modification of FMM for KMC. We show that this overcomes the fundamental issues described in \cite{Li2018}. With our modified FMM algorithm the cost per KMC step grows linearly in the number of charges (and not quadratically as claimed in \cite{Li2018}) and accurate electrostatic interactions can be included in large-scale KMC simulations. The key observation is that - since FMM describes the long-range contribution as a continuous field - the change in the electrostatic potential energy can be evaluated at a cost of $\mathcal{O}(1)$ (i.e. independent of the particle number) for each proposed hopping event. As there are $\mathcal{O}(N)$ potential hopping events per KMC step, the total computational complexity of the propensity calculation is $\mathcal{O}(N)$. Updating the FMM field after one hop is accepted carries an additional cost of $\mathcal{O}(N)$, resulting in a total computational complexity per KMC step which scales linearly in the number of charges.

While not the topic of this paper, we remark that it is also possible to improve the computational complexity of \textit{standard} Monte Carlo (MC) by similar methods. As will be argued at the end of this paper, we believe that changes to the electrostatic energy for each \textit{individual at\-tempt\-ed MC move} can be calculated at a computational cost $\mathcal{O}(\log(N))$ with a suitably modified version of FMM.

To simulate large physical systems, an efficient, parallel implementation of the algorithm is important to obtain meaningful results in a reasonable time. Easy integration into existing simulation packages and workflows can be achieved by providing a minimal yet flexible user-interface. With the recent diversification of the hardware landscape, the code should be performance portable and run on different chip architectures, including, for example, traditional CPUs and GPUs. The implementation described in this paper is based on the performance portable framework first introduced in \cite{Saunders2018}. By providing a Python interface and using code generation techniques, the code is fast, yet allows the user to express their algorithms at a high abstraction level.

For an idealised setup we find that our FMM-KMC algorithm can be used to carry out simulations with exact electrostatics on problems with $10^6$ charges in $0.14$s per KMC step when running on a parallel computer with $8192$ cores. In a physically realistic configuration the hopping processes of 20412 particles in a $\alpha$-NPD problem doped with F6TCNNQ at a concentration of $2\%$ could be simulated at a rate of $0.35\text{s}$ per KMC step on a single 12-core Skylake CPU. 
\pparagraph{Structure}
This paper is organised as follows:
After reviewing the key concepts of KMC and FMM in Section \ref{sec:methods} we describe our adaptation of the FMM algorithm for KMC simulations in Section \ref{sec:FMMforKMC} and review related work in Section \ref{sec:literature_review}. An efficient implementation of our method based on the performance portable framework in \cite{Saunders2018} is described in Section \ref{sec:implementation}, where we discuss the user-interface in detail. Numerical results which demonstrate the accuracy, computational efficiency and parallel scalability of the algorithm for idealised model systems and a physically relevant setup are presented in Section \ref{sec:results}. Finally, we conclude and discuss possible future directions of our work in Section \ref{sec:conclusions}. Some more technical aspects are relegated to the appendices. The standard FMM algorithm is written down in \secapp\ref{sec:FMM_algorithm_details} and the correction term for charge distributions with a non-vanishing dipole-moment is derived in \secapp\ref{sec:dipole_correction_details}. An improved user interface for proposing moves, which is optimised for efficiency, is described in \secapp\ref{sec:improved_proposal_interface}. Previously, we reported on the performance of Ewald-based long range electrostatics in the same code base \cite{Saunders2017a}. To complement this work we discuss the performance and scalability of the standard FMM implementation in \secapp\ref{sec:results_parallelFMM}.
\section{Review of Methods}\label{sec:methods}
To put our new algorithm into context and establish necessary notation, we first review the KMC method and the standard FMM algorithm.
\subsection{Kinetic Monte Carlo}
Molecular Dynamics (MD) and Monte Carlo (MC) are the standard computational tools for predicting the properties of physical and chemical systems from first principles (see e.g. \cite{Frenkel2001,Allen1989}). Typically it is assumed that the molecular constituents interact via phenomenological classical potentials; for charged systems this includes long range electrostatic interactions. While MC can be used to study systems in equilibrium by sampling from the steady state distribution, MD allows the simulation of dynamical processes such as time dependent charge propagation in batteries and solar cells. In solid state systems at moderate temperatures and pressures there are often two types of processes which occur at very different time scales: fast oscillations around local minima of the energy landscape, which are separated by large energy barriers, and much slower transitions between those minima. In a system with these properties MD is highly inefficient for extracting quantities such as charge mobilities and voltage characteristics. This is because the MD timestep needs to be small enough to resolve the fast oscillations, yet the trajectories have to be sufficiently long to include dynamically relevant transitions between local minima. In fact, the rapid oscillations do not contain interesting physical information on charge transport processes, and should be integrated out. Kinetic Monte Carlo (KMC) \cite{Young1966,Bortz1975,Gillespie1976,Gillespie1977} overcomes this problem by treating each of the local minima as an independent configuration or state $S_{\stateidxi}$ (here and in the following indices $\stateidxi,\stateidxj,\dots$ are used to label states; particles are indexed with Roman letters $i,j,\dots$ and Greek letters $\alpha,\beta,\dots$ are used for cells in the computational grid). The dynamics are approximated by probabilistic transitions between those configurations. The generated probabilistic trajectory is equivalent to snapshots of the full MD simulation at discrete times. For example, free charge carriers in a crystal at room temperature are bound to specific local sites, and the states correspond to particular distributions of the particles, such that every site is either empty or occupied. KMC assumes that there is a fixed rate $r_{\stateidxi\stateidxj}$ for a configuration to transition from state $S_{\stateidxi}$ to state $S_{\stateidxj}$ in a given time, i.e. each transition is modelled as a Poisson process. The rates $r_{\stateidxi\stateidxj}$ are known as ``propensities'' in the KMC literature. This is a valid approximation if we assume that - compared to the state-transition time scale - the fast oscillations around the local equilibria occur so rapidly that the particles ``forget'' their previous history, and the transition probability is the same at each point in time. The propensities are inputs to the KMC algorithm and it typically assumed that $r_{\stateidxi\stateidxj}$ depends on the energy difference between states $S_{\stateidxi}$ and $S_{\stateidxj}$. Since transitions between states are modelled by a Poisson process, the probability distribution function for the time of the first escape from the state $S_{\stateidxi}$ is
\begin{equation}
  \pi_{\stateidxi}(t) = Q_{\stateidxi} e^{-Q_{\stateidxi} t} \qquad\text{where $Q_{\stateidxi} := \sum_{\stateidxj=1}^{\Nstates} r_{\stateidxi\stateidxj}$}.
\label{eqn:p_transitiontime}
\end{equation}
Here $\Nstates$ is the total number of states. This information is used to work out the physical time interval between subsequent snapshots. The following two steps occur at each transition between states:
\begin{enumerate}
  \item Starting from state $S^{(n)}=S_{\stateidxi}$, pick a new state $S^{(n+1)}=S_{\stateidxj}$ such that the the probability of transitioning from state $S_{\stateidxi}$ to $S_{\stateidxj}$ is proportional to $r_{\stateidxi\stateidxj}$.
  \item Increment the current time by drawing from the distribution in Eq. \eqref{eqn:p_transitiontime}. This can be achieved by choosing a uniform random number $\xi\in(0,1]$ and setting the time increment $\Delta t=-Q_{\stateidxi}^{-1}\log(\xi)$.
\end{enumerate}
Note that while the method is written down for a finite number $\Nstates$ of states in Algorithm \ref{alg:KMC}, it also works for systems with an infinite number of configurations, if it is assumed that in each step of the algorithm only a finite number of other states can be reached. This is often a sensible assumption since particles can only hop to nearby sites.
\begin{figure}
\begin{center}
\includegraphics[width=0.5\linewidth]{\figdir/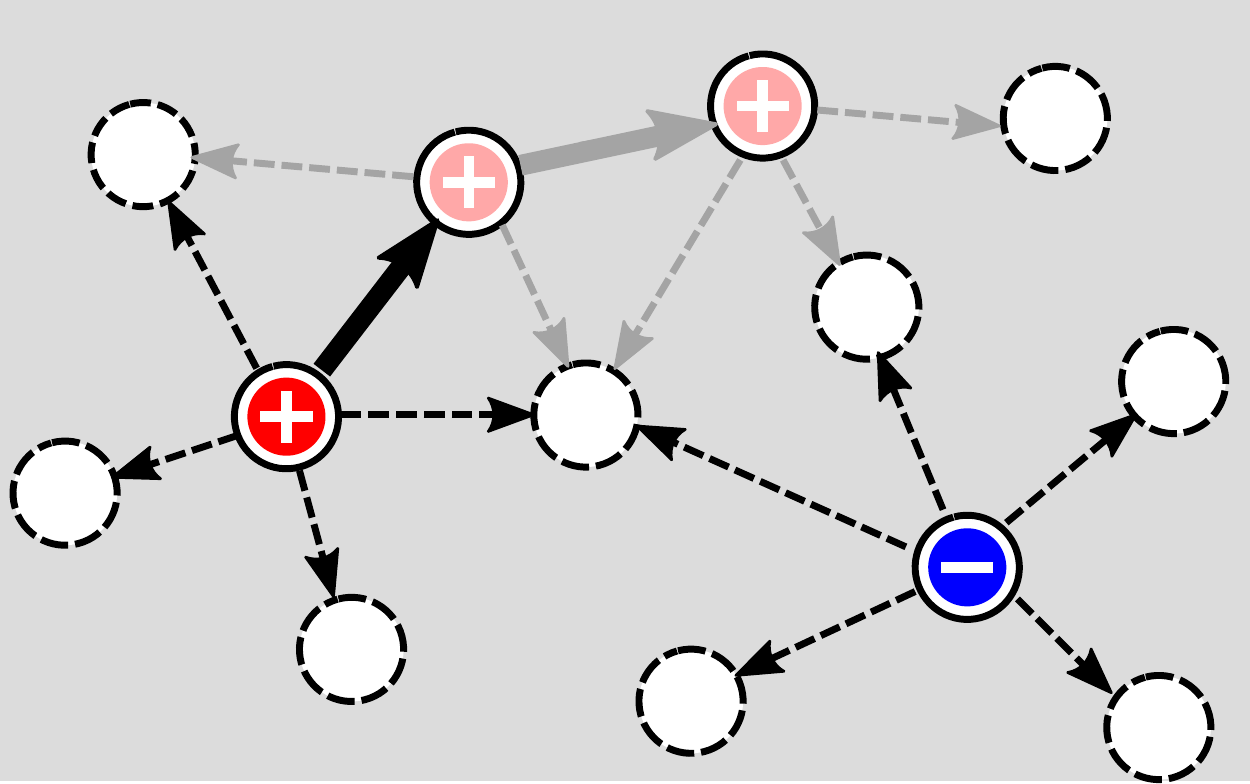}
\caption{Schematic sketch of one KMC step, which consists of calculation of the propensities for all possible hops (dashed arrows) and moving one particle to a new site after accepting a particular hop (solid arrows). Subsequent steps are shown as gray arrows.}
\label{fig:fmm_step_schematic}
\end{center}
\end{figure}

For a particular problem the propensities $r_{\stateidxi\stateidxj}$ are an input for the algorithm and need to be calculated, for example by working out the Boltzmann factors of different configurations. Crucially, this calculation requires knowledge of the change $\Delta U_{\stateidxi\stateidxj}=U_{\stateidxj}-U_{\stateidxi}$ in system energy induced by the hop. Including the contribution of the electrostatic interaction to this energy difference is very expensive and requires efficient algorithms.
\begin{algorithm}
\caption{Kinetic Monte Carlo (KMC) method for generating snapshots $S^{(0)},S^{(1)},\dots,S^{(n)}$ of the system dynamics.}
\label{alg:KMC}
\begin{algorithmic}[1]
\STATE{Pick initial state $S^{(0)}$, set $t=0$, $n=0$}
\WHILE{$t<T$}
\STATE{Set $i$ such that $S^{(n)}=S_{\stateidxi}$}
\FORALL{states $\stateidxj=1,\dots,\mathfrak{N}$}
\STATE{Calculate difference $\Delta U_{\stateidxi\stateidxj}=U_{\stateidxj}-U_{\stateidxi}$}
\STATE{Derive propensities $r_{\stateidxi\stateidxj}=r(\Delta U_{\stateidxi\stateidxj})$}
\STATE{Calculate $R_{\stateidxi\stateidxj} = \sum_{\stateidxk=1}^{\stateidxj} r_{\stateidxi\stateidxk}$, $Q_{\stateidxi}=R_{\stateidxi \Nstates}$}
\ENDFOR
\STATE{Draw a uniform random number $\zeta\in(0,1]$}
\STATE{Set $S^{(n+1)}=S_{\stateidxj}$ with $R_{\stateidxi,\stateidxj-1}<Q_{\stateidxi}\zeta \le R_{\stateidxi\stateidxj}$}
\STATE{Draw another uniform number $\xi\in(0,1]$}
\STATE{Calculate time increment $\Delta t=-Q_{\stateidxi}^{-1}\log(\xi)$}
\STATE{Set $n\mapsto n+1$, $t\mapsto t + \Delta t$}
\ENDWHILE
\end{algorithmic}
\end{algorithm}
\subsection{The Fast Multipole Method}\label{sec:FMM}
\begin{figure}
\begin{center}
\includegraphics[width=0.5\linewidth]{\figdir/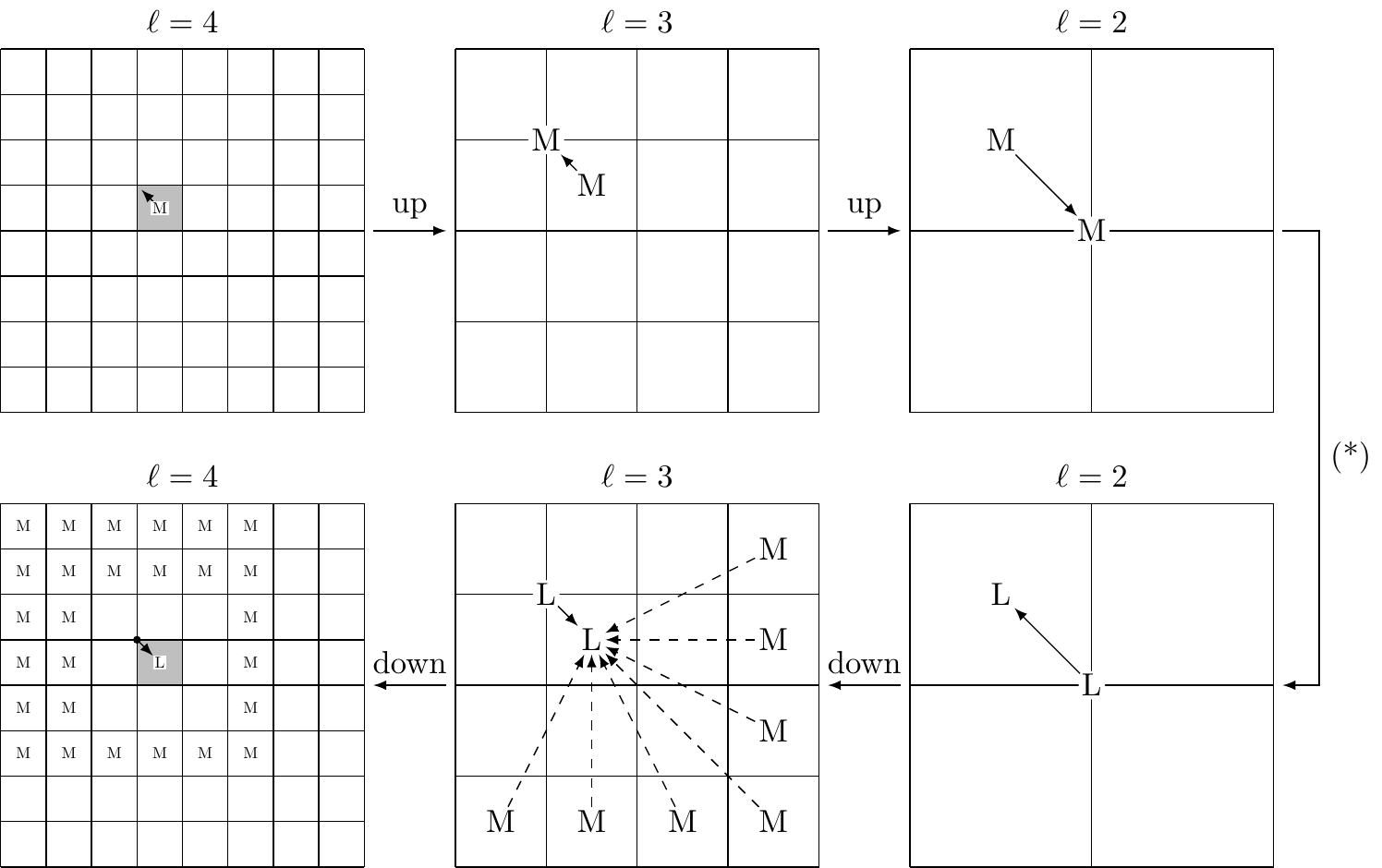}
\caption{Overview of the FMM method, see Algorithm \ref{alg:fmm} in \secapp\ref{sec:FMM_algorithm_details} for details. The translation of multipole expansions $\Phi_{\ell,\cellidxi}$ in the upward pass is shown in the first row. The second row shows the conversion of multipole expansions in the interaction list into local expansions $\Psi_{\ell,\cellidxi}$, which are represented on the next finer level. The asterisk on the right hand side of the figure stands for any additional operations on the coarsest level to account for the boundary conditions (see Section \ref{sec:boundary_conditions}).}
\label{fig:fmm_schematic}
\end{center}
\end{figure}

To allow an in-depth understanding of the proposed new algorithm for electrostatic interactions in KMC, we first describe the classical Fast Multipole Method (FMM) introduced in \cite{Greengard1987}, before discussing its adaptation in Section \ref{sec:FMMforKMC}. For further technical details we refer the reader to the original literature \cite{Greengard1987,Greengard1988,Greengard1997}. In three dimensions the FMM algorithm uses a hierarchical grid with $L$ levels for the computational domain $\Omega$ (which is assumed to be a equilateral cube of width $a$) such that the number of cells on each level $\ell$ is $M_\ell = 8^{\ell-1}$ for $\ell=1,\dots,L$. The number of cells on the finest level is $M=M_L$, and typically $L$ is chosen such that there there are $\mathcal{O}(1)$ particles in each fine level cell. Each cell on level $\ell=1,\dots,L-1$ is subdivided into 8 child-cells on the next-finer level; conversely each cell on level $\ell=2,\dots,L$ has a unique parent cell. The Fast Multipole Algorithm now computes the electrostatic potential by splitting it into two contributions. First, the long range part is calculated by working out the multipole expansion of all charges in a fine level cell, followed by an upward- and downward traversal of the grid hierarchy (see Fig. \ref{fig:fmm_schematic}). In the upward pass of the algorithm the multipole expansions around the centre of a cell are recursively combined and converted to multipole expansions around the centre of the parent cell, obtaining a single multipole expansion around the centre of the computational domain on the coarsest level $\ell=1$. In the downward pass the multipole expansions on each level are transformed into local expansions around the centre of a cell. Those are then recursively combined into local expansions in the child cells. By only considering the contribution from multipole expansions in a fixed number of well-separated cells on each level, the contribution from distant charges are resolved at the appropriate level of accuracy, while including the contribution from closer charges in finer levels.
The $p$-term multipole expansions $\Phi$ and the local expansions $\Psi$ which play a central role in the FMM algorithm can be expressed in terms of the spherical harmonics $Y_n^m(\theta,\phi)$, i.e.
\begin{xalignat}{2}
    \Phi(r,\theta,\phi) &= \sum_{n=0}^{p}\sum_{m=-n}^{+n} M_n^m r^{-(n+1)}Y_n^m(\theta,\phi),
& \Psi(r,\theta,\phi) &= \sum_{n=0}^{p}\sum_{m=-n}^{+n} L_n^m r^{n}Y_n^m(\theta,\phi).\label{eqn:multipole_local_expansion}
\end{xalignat}
where $(r,\theta,\phi)$ are spherical coordinates relative to a suitable origin.

Overall it can be shown \cite{Greengard1987,Greengard1988,Greengard1997} that at leading order the computational cost of the method is
\begin{equation*}
  \text{Cost}_{\text{FMM}} = C p^4 N + \dots
\end{equation*}
where $p$ is the order where the multipole and local expansions in Eq. \eqref{eqn:multipole_local_expansion} are truncated. To bound the error by some $\epsilon$, the value of $p$ needs to be chosen such that $p = \mathcal{O}(\log_2 \epsilon)$.

The calculation of the local expansions on the finest level is written down explicitly in Algorithm \ref{alg:fmm} in \secapp\ref{sec:FMM_algorithm_details} and requires the following definitions, which will be used in the subsequent discussion of the FMM method for KMC simulations:
\begin{description}
    \item[$\Phi_{\ell,\cellidxi}$]{the $p$-term multipole expansion (see Eq. \eqref{eqn:multipole_local_expansion}) about the centre of cell $\cellidxi$ on level $\ell$ that describes the potential induced by all charges contained in cell $\cellidxi$.}
    \item[$\Psi_{\ell,\cellidxi}$]{the $p$-term local expansion (see Eq. \eqref{eqn:multipole_local_expansion}) about the centre of cell $\cellidxi$ on level $\ell$ that describes the potential induced by all charges outside the cell $\cellidxi$ and its 26 nearest neighbours.}
    \item[$\mathcal{T}_{\operatorname{MM}}$]{the linear operator translating multipole moments to multipole moments around a different origin.}
    \item[$\mathcal{T}_{\operatorname{ML}}$]{the linear operator converting multipole moments to coefficients of the local expansion.}
    \item[$\mathcal{T}_{\operatorname{LL}}$]{the linear operator translating local expansion coefficients to local expansion coefficients around a different origin.}
\end{description}
\begin{figure}
    \begin{center}
        \includegraphics[width=0.45\linewidth]{\figdir/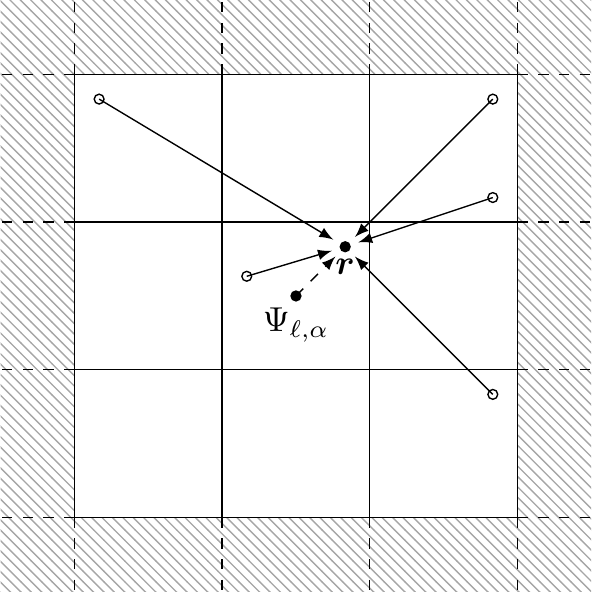}
        \caption{The potential field at $\vec{r}$ is given by the sum of (1) the evaluation of $\Psi_{\ell,\cellidxi}$ at $\vec{r}$ (long range) and (2) direct interactions with charges in the white region (short range).
        Each solid arrow illustrates a direct interaction with a charge represented by an empty circle.}
        \label{fig:fmm_point_eval}
    \end{center}
\end{figure}
Finally, the short range contribution of the electrostatic potential is obtained by calculating the field generated by charges in neighbouring cells directly. Fig. \ref{fig:fmm_point_eval} illustrates how the total calculation is split up into the long- and short-range contributions discussed above. The algorithm for calculating the total electrostatic energy from the local expansions $\Psi_{L,\alpha}$ on the finest level is given in Algorithm \ref{alg:fmm_evaluation} in \secapp\ref{sec:FMM_algorithm_details}.
\subsubsection{Boundary conditions}\label{sec:boundary_conditions}
So far we assumed free-space boundary conditions, i.e. we consider a finite charge distribution contained inside an unbounded physical domain. In this case the interaction list is empty on the two coarsest levels, and we can set $\Psi_{1,1}=\overline{\Psi}_{1,1}=\Psi_{2,\cellidxi} = \overline{\Psi}_{2,\cellidxi}=0$ or equivalently skip those two levels in the downward pass.

\begin{figure}
\begin{center}
\includegraphics[width=0.35\linewidth]{\figdir/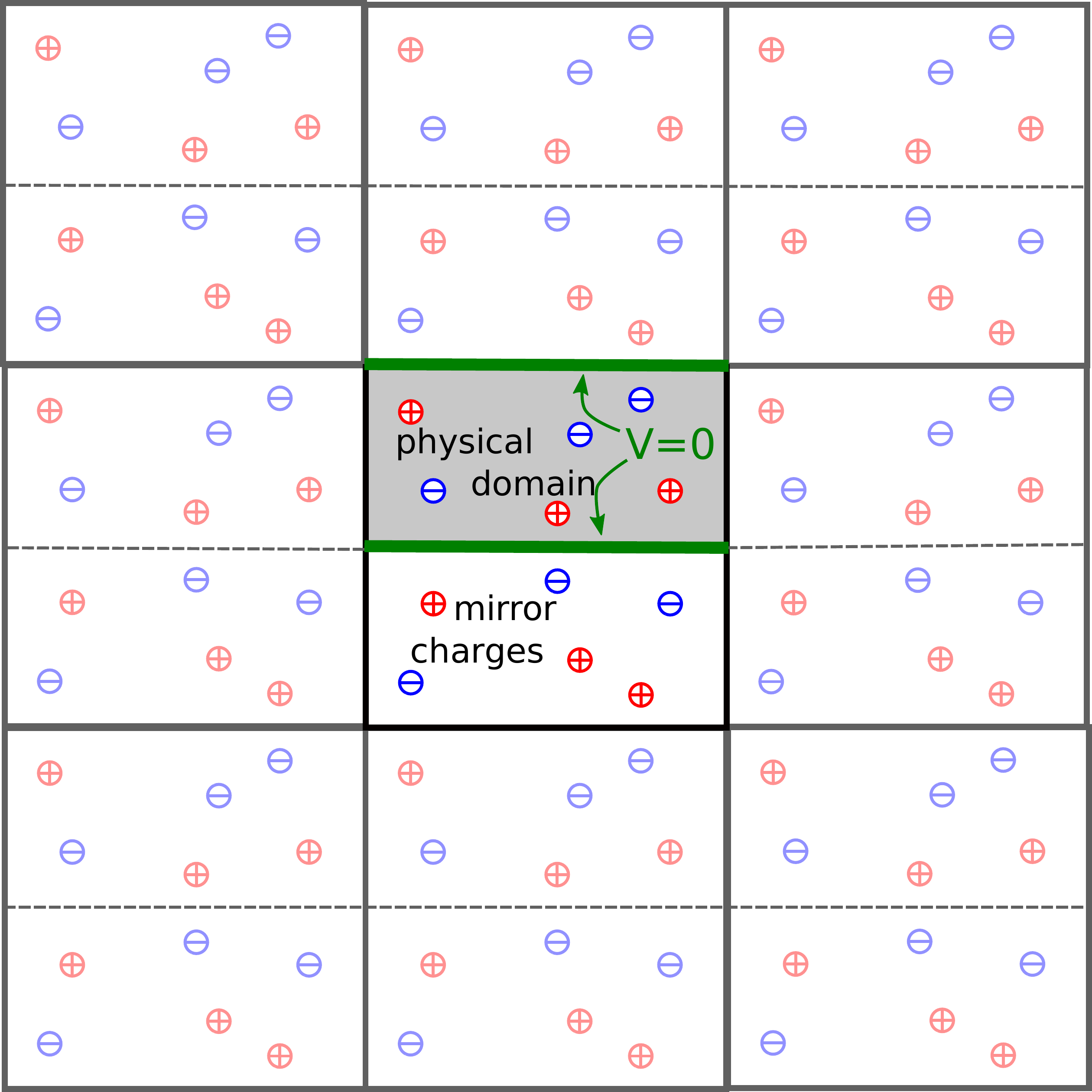}
\caption{Computational domain used for zero-potential Dirichlet boundary conditions $V=0$ at the top and bottom of the device, which is highlighted in green. After duplicating the charges in the physical domain (gray background) with mirror charges, the entire computational domain (black box) is replicated periodically.}
\label{fig:dirichlet_bc_domain}
\end{center}
\end{figure}

When simulating large physical systems, however, the computational domain is typically replicated in one or several space dimensions to avoid spurious surface effects from the finite computational domain. The FMM algorithm is readily modified to account for this, as we discuss in the following for two important cases. Note that care has to be taken if the system has a net charge or a non-zero dipole moment - in those cases the lowest-order sums over periodic copies is conditionally convergent and need to be fixed using physical conditions, see appendix 4.1 of \cite{Greengard1987} and \secapp\ref{sec:dipole_correction_details} of this paper.
\pparagraph{Periodic}
In the simplest case the computational domain is replicated periodically. To account for this, the operator $\mathcal{T}_{\operatorname{ML}}$ has to be modified on the coarsest level. On this level the local expansion receives contributions from the multipole expansions in an infinite number of well-separated periodic copies. In \cite{Amisaki2000} the contributions from those copies are summed with an Ewald-like method and it is shown that they can be accounted for by simply replacing the spherical harmonics $Y_\ell^m$ which appear in the linear operator $\mathcal{T}_{\operatorname{ML}}$ by an infinite sum $R_\ell^m$. In other words, the local expansion $\Psi_{1,1}$ on the coarsest level can be obtained from the multipole expansion $\Phi_{1,1}$ on the same level through multiplication by a known linear operator: $\Psi_{1,1}=\mathcal{R}\Phi_{1,1}$. The sum $R_\ell^m$ and the linear operator $\mathcal{R}$ can be calculated once at the beginning of the simulation.
\pparagraph{Dirichlet}
In some cases it is desirable to apply homogeneous Dirichlet boundary conditions $\phi(\vec{x})=0$ on some or all boundaries of the computational domain. This allows the inclusion of an external electric field which is present for example in batteries. As discussed in Appendix 4.2 of \cite{Greengard1987}, this can be achieved by adding appropriate virtual mirror charges, effectively replacing the infinite grid of periodic copies by suitably modified reflected and inverted copies of the primary charge distribution. Apart from adjusting the value of the sum $R_\ell^m$ this will require no further modifications of the algorithm. An alternative approach, which we pursue here (and which is also described in \cite{Casalegno2010}), is to simply extend the computational domain with the first mirror charge image and replicating the extended domain periodically, as shown in Fig. \ref{fig:dirichlet_bc_domain}. We then apply the above algorithm for periodic boundary conditions to this extended domain. In the physical applications we consider here the potential is fixed at the top and bottom of the domain, which is assumed to be periodic in the other two space dimensions. This will only lead to a potential loss of performance by at most a factor two from doubling the number of charges in the system.

\subsubsection{Dipole correction}\label{sec:dipole_correction}
In the case of non-trivial boundary conditions, the contribution of the dipole terms to the infinite sum which determines the local expansion is conditionally convergent. As discussed in \cite{Greengard1987}, the value of the sum needs to be fixed by physical considerations. For example, following section 4.1 of \cite{Greengard1987}, one could require that for a configuration which consists of a pure dipole pointing in the $z$ direction the difference $\Delta\phi=\phi(\vec{r}_{a/2})-\phi(-\vec{r}_{a/2})$ in the potential between the points $\vec{r}_{a/2}=(0,0,a/2)$ and $-\vec{r}_{a/2}$ vanishes. This is not the case for the treatment in \cite{Amisaki2000}, which induces a constant electric field in the $z$-direction in the presence of a dipole; physically this corresponds to a non-zero surface charge at infinity.
In our calculations we choose to require $\Delta\phi=0$, i.e. no surface charge at infinity. As explained in \secapp\ref{sec:dipole_correction_details}, this can be achieved by adding a compensating external electric field $\vec{E}=\frac{4\pi}{3}\vec{p}$ where $\vec{p}$ is the dipole moment in the simulation cell. In practice this amounts to modifying the local expansion coefficients $L_1^m$ defined in Eq. \eqref{eqn:multipole_local_expansion} on the coarsest level.
\section{FMM for KMC}\label{sec:FMMforKMC}
We now describe how the FMM algorithm can be used to compute electrostatic energy differences required for the calculation of propensities in KMC simulations. In contrast to a naive approach, where the classical FMM algorithm described in Section \ref{sec:FMM} is simply used as a black-box solver to calculate the charge distribution after each proposed move, here the multipole method is closely integrated with the KMC update. For simplicity, we initially consider free-space boundary conditions, before discussing the modifications which are required to adapt the algorithm to periodic- and Dirichlet- boundary conditions in Section \ref{sec:fmm_for_kmc_periodic}. As we will show, using the correctly modified FMM results in a total computational complexity of $\mathcal{O}(p^2N)$ per KMC step (the complexity is $\mathcal{O}(p^4N)$ for non-trivial boundary conditions). In each step, the electrostatic calculation can be split into two parts:
\begin{itemize}
\item \textbf{Propose moves}: calculate the change in electrostatic energy $\Delta U_{\stateidxi\stateidxj}=U_\stateidxj-U_\stateidxi$ for all potential moves (line 5 of Algorithm \ref{alg:KMC})
  \item \textbf{Accept move}: update the electrostatic potential, i.e. the local expansions $\Psi_{L,\cellidxi}$, once a move has been accepted (line 10 of Algorithm \ref{alg:KMC}).
\end{itemize}

We further assume that the standard FMM algorithm in Algorithm \ref{alg:fmm} has been used to calculate an expansion of the local field $\Psi_{L,\cellidxi}$ from long-range contributions before the first KMC step. Since the number of KMC moves is very large (compared to $p^2$), this $\mathcal{O}(p^4N)$ startup cost can be safely neglected. In the following the calculation of electrostatic energy differences in the proposal stage and the update of FMM data structures are discussed separately.
\pparagraph{Propose moves}
Let $\vec{r}'$ be the proposed new position of a particle with charge $q$ currently located at position $\vec{r}$ in cell $\cellidxi$ on the finest level $L$ of the FMM grid hierarchy. Further denote the 26 direct neighbours of a cell $\cellidxi$ as $\mathcal{N}_b(\cellidxi)$ and define $\ooverline{\cellidxi} = \cellidxi\cup\mathcal{N}_b({\cellidxi})$. Assuming that the new position is in cell ${\cellidxi}'$ (which might be identical to $\cellidxi$), the total change in the electrostatic energy due to the move $\vec{r}'\leftarrow \vec{r}$ is given by
\begin{equation}
  \begin{aligned}
  \Delta U_{\vec{r}'\leftarrow \vec{r}} &=
  q\left(\Psi_{L,{\cellidxi}'}(\vec{r}') + \sum_{\vec{r}^{(\partidxi)}\in {\ooverline{\cellidxi'}}} \frac{q^{(\partidxi)}}{\left|\vec{r}'-\vec{r}^{(\partidxi)}\right|}\right)
  - q\left(\Psi_{L,\cellidxi}(\vec{r}) + \sum_{\substack{\vec{r}^{(\partidxi)}\in\ooverline{\cellidxi}\\\vec{r}^{(\partidxi)}\ne \vec{r}}} \frac{q^{(\partidxi)}}{\left|\vec{r}-\vec{r}^{(\partidxi)}\right|}\right)
  -\frac{q^2}{|\vec{r}'-\vec{r}|}.
  \end{aligned}\label{eqn:proposed_energy}
\end{equation}
The first two brackets in Eq. \eqref{eqn:proposed_energy} describe the difference in electrostatic energy of the particle at the new and old position, split into a long range part (given by the local expansions $\Psi_{L,\cellidxi}$ and $\Psi_{L,\cellidxi'}$) and direct interactions with all particles in the cell which contains the particle and its direct neighbours. Note that the terms in the first bracket, which describes the potential energy after the move, contain the potential induced by the particle at position $\vec{r}$ before the move. This contribution is contained either implicitly in $\Psi_{L,\cellidxi'}$, if the cells $\cellidxi$ and $\cellidxi'$ are well separated, or included in the direct contribution if this is not the case, because one of the $\vec{r}^{(\partidxi)}$ is identical to $\vec{r}$. Clearly this is incorrect since the particle has moved to $\vec{r}'$ in this proposal and is no longer at $\vec{r}$. This is fixed by removing the spurious self-interaction in the final term of Eq. \eqref{eqn:proposed_energy}.

Since the local expansion $\Psi_{L,\cellidxi}$ defined in Eq. \eqref{eqn:multipole_local_expansion} consists of $\mathcal{O}(p^2)$ terms, and each cell contains $\overline{N}_{\text{local}}=\mathcal{O}(1)$ particles on average, the total cost per proposal is
\begin{equation}
  \text{Cost}_{\text{propose}}^{(\text{free})}=C p^2 + 27\overline{N}_{\text{local}}=\mathcal{O}(p^2),\label{eqn:cost_propose}
\end{equation}
independent of the total number of charges.
\pparagraph{Accept move}
Once a move $\vec{r}'\leftarrow\vec{r}$ has been accepted, the local expansion $\Psi_{L,\cellidxj}$ on the finest level has to be updated in all cells $\cellidxj$ to account for this. Naively, this could be done by re-calculating the entire field using Algorithm \ref{alg:fmm}, at a cost of $\mathcal{O}(p^4N)$. However, since the change in the charge distribution is only very small, the change in $\Psi_{L,\cellidxj}$ can be computed much more efficiently by subtracting the contribution of the charge at the original position $\vec{r}$ and adding it back on at the new position $\vec{r}'$. For this, loop over all cells $\cellidxj$ on the finest level. If cell $\cellidxj$ is well separated from cell $\cellidxi$ (i.e. $\cellidxj\not\in \cellidxi\cup\mathcal{N}_b(\cellidxi))$, add the local expansion around the centre of cell $\cellidxi$ which is induced by a monopole with charge $-q$ at position $r$ to $\Psi_{L,\cellidxi}$, using a multipole-to-local translation $\mathcal{T}_{\text{ML}}$. Similarly, if $\cellidxj$ is well separated from $\cellidxi'$, add the local expansion induced by a monopole of charge $+q$ at the new position $\vec{r}'$. Since the local expansion contains $\mathcal{O}(p^2)$ terms and the total number of cells on the finest level is $\mathcal{O}(N)$, this reduces the cost of the accept step to
\begin{equation}
  \text{Cost}_{\text{accept}}^{(\text{free})} = C'p^2 N = \mathcal{O}(p^2N).
  \label{eqn:cost_accept}
\end{equation}
For higher degrees $p$ this leads to significant savings of a factor $p^2$ relative to the naive re-calculation with Algorithm \ref{alg:fmm}.

The above method is readily extended to non-trivial boundary conditions introduced in Section \ref{sec:boundary_conditions} as follows.
\subsection{Boundary conditions}\label{sec:fmm_for_kmc_periodic}
\begin{figure}
    \centering
    \includegraphics[width=0.25\linewidth]{\figdir/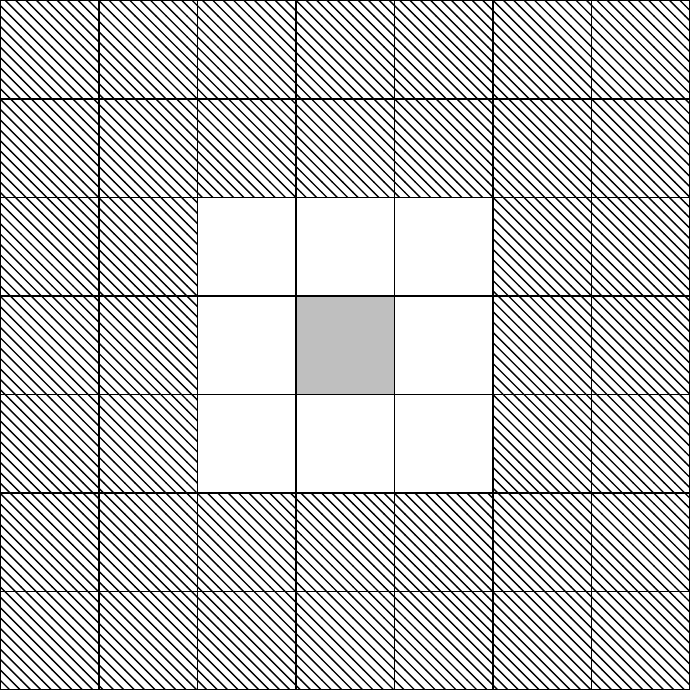}
    \caption{Sketch of $V^{(\sr)}$ and $V^{(\lr)}$ defined in Section \ref{sec:fmm_for_kmc_periodic}. $V^{(\lr)}$ contains the outer hatched cells and $V^{(\sr)}$ consists of the primary image (solid grey) and the surrounding nearest neighbours (empty cells).}
    \label{fig:sr_lr_split}
\end{figure}

\pparagraph{Periodic boundary conditions}
When periodic boundary conditions are applied, the simulation cell is surrounded by an infinite lattice of periodic images of the domain $\Omega$ indexed by integer valued offsets $\vec{\nu} \in \mathbb{Z}^3$ (where $\vec{\nu} = \vec{0}$ corresponds to the primary image).
This infinite lattice is split into two disjoint sets $\mathbb{Z}^3=V^{(\sr)}\cup V^{(\lr)}$ as shown in Fig. \ref{fig:sr_lr_split}. The finite set
\begin{equation*}
V^{(\sr)}=\{\vec{\nu}:|\nu_k|\le 1\;\text{for all}\;k=1,2,3\}
\end{equation*}
contains the primary image and the surrounding 26 nearest neighbours and
\begin{equation*}
  V^{(\lr)}=\{\vec{\nu}:|\nu_k|>1\;\text{for at least one}\;k=1,2,3\}
\end{equation*}
consists of all other periodic copies.
Due to linearity, the system energy is given by a contribution from periodic images in $V^{(\sr)}$ plus the contribution from images in $V^{(\lr)}$.
For a proposed move we consider the contribution to system energy from each of these two sets separately and sum them to obtain the total change in system energy.

We refer to the contribution to the system energy from images in $V^{(\lr)}$ as the \textit{far-field} component and the contribution from images in $V^{(\sr)}$ as the \textit{near-field} component. As before, let $\vec{r}'$ be the proposed new position of a particle with charge $q$ currently located at $\vec{r}$ in cell $\cellidxi$ on the finest level $L$ of the FMM grid hierarchy and assume that the new position $\vec{r}'$ is in cell $\cellidxi'$.
\pparagraph{Near-field component}
To compute the near-field component we simply extend the method described for free-space boundary conditions to include all images in $V^{(\sr)}$; instead of moving a single particle, we also move its 26 copies in $V^{(\sr)}$ when proposing a move.

While this could be achieved by working with all charges in the 27 cells in $V^{(\sr)}$ and employing free-space boundary conditions, in practice it is more efficient to work with the primary image only and implicitly include the 26 copies. To achieve this, first the FMM algorithm in Algorithm \ref{alg:fmm} is modified such that at the end of the downward pass a local expansion $\Psi_{L,\cellidxi}$ represents the potential induced by all charges in $V^{(\sr)} \setminus \mathcal{N}_b(\cellidxi)$. This can be realised by setting $\Psi_{1,1}=\overline{\Psi}_{1,1}=\Psi_{2,\cellidxi} = \overline{\Psi}_{2,\cellidxi}=0$ at the beginning of the downward pass.
Secondly, the interaction lists used in the downward pass are not truncated at the boundary of the primary domain and instead are wrapped around the boundary to account for the neighbours in $V^{(\sr)}$. In a similar manner the neighbour cells $\mathcal{N}_b(\cellidxi)$ of a boundary cell $\cellidxi$ include cells over the boundary in $V^{(\sr)}$. With these modifications the contribution to the system energy from charges in $V^{(\sr)}$ is
\begin{equation*}
  \begin{aligned}
    U^{(\sr)} &= \frac{1}{2} \sum_{\partidxi=1}^N  q^{(\partidxi)} \omega^{(\partidxi)}\qquad\text{with}\\
    \omega^{(\partidxi)} &= \Psi_{L,\cellidxi[\partidxi]}(\vec{r}^{(\partidxi)}) + \sum_{\substack{\vec{r}^{(\partidxj)}\in \ooverline{\cellidxi[\partidxi]}^*\\\partidxi\neq\partidxj}} \frac{q^{(\partidxj)}}{\left|\vec{r}^{(\partidxi)}-\vec{r}^{(\partidxj)}\right|}
    \end{aligned}
\end{equation*}
where $\cellidxi[\partidxi]$ is the index of the fine-level cell which contains particle $\partidxi$. The asterisk ($*$) on $\ooverline{\cellidxi[\partidxi]}^*$ indicates that neighbours are wrapped periodically across the boundary of the computational domain, as discussed above. The factor $\frac{1}{2}$ is required to avoid double-counting in the total energy.

Consequently, the change in the near-field system energy for the proposed move $\vec{r}'\leftarrow \vec{r}$ is
\begin{equation}
  \begin{aligned}
      \Delta U^{(\sr)}_{\vec{r}'\leftarrow \vec{r}} &=
      q\left(\Psi_{L,\cellidxi'}(\vec{r}') + \sum_{\vec{r}^{(\partidxi)}\in{\ooverline{\cellidxi'}^*}} \frac{q^{(\partidxi)}}{\left|\vec{r}'-\vec{r}^{(\partidxi)}\right|}\right)
      - q\left(\Psi_{L,\cellidxi}(\vec{r}) + \sum_{\substack{\vec{r}^{(\partidxi)}\in\ooverline{\cellidxi}^*\\\vec{r}^{(\partidxi)}\ne \vec{r}}} \frac{q^{(\partidxi)}}{\left|\vec{r}-\vec{r}^{(\partidxi)}\right|}\right)\\
      &\quad-\;\;q^2 \sum_{\vec{\nu} \in V^{(\sr)} } \frac{1}{|\vec{r}'-(\vec{r}+a\vec{\nu})|} 
      +q^2 \sum_{\substack{\vec{\nu} \in V^{(\sr)}\\ \vec{\nu} \ne \vec{0} }} \frac{1}{|\vec{r}'-(\vec{r}'+a\vec{\nu})|}.
  \end{aligned}\label{eqn:proposed_energy_sr}
\end{equation}
Eq. \eqref{eqn:proposed_energy_sr} is identical to Eq. \eqref{eqn:proposed_energy} except for the final two terms. The penultimate term is a generalisation of the self-energy correction in Eq. \eqref{eqn:proposed_energy}, which also includes the periodic copies of $\vec{r}$. The final term accounts for the fact that all periodic copies of the particle are moved to new positions $\vec{r'}+a\vec{\nu}$ with $\vec{0}\ne\vec{\nu}\in V^{(\sr)}$, and those charges contribute to the energy of the particle at $\vec{r}'$. The sum is independent of $\vec{r}'$ and it is readily evaluated to
\begin{equation*}
  \begin{aligned}
    \sum_{\substack{\vec{\nu} \in V^{(\sr)}\\ \vec{\nu} \ne \vec{0} }} \frac{1}{|\vec{r}'-(\vec{r}'+a\vec{\nu})|}
    &= \frac{1}{a} \sum_{\substack{\vec{\nu}\in V^{(\sr)}\\ \vec{\nu}\ne \vec{0}}} |\vec{\nu}|^{-1}
      = \frac{1}{a} \left( 6+\frac{8}{\sqrt{3}}+\frac{12}{\sqrt{2}} \right)
    \approx \frac{19.104}{a} 
  \end{aligned}
  \end{equation*}
\pparagraph{Far-field component}
We compute the potential induced by charges in $V^{(\lr)}$ by following the approach used for the standard FMM algorithm \cite{Amisaki2000}.
In the setup phase of the simulation the multipole expansion $\Phi_{1,1}$ is computed directly from the initial configuration. The $(n,m)^\text{th}$ multipole coefficient is
\begin{equation*}
    K_n^m = \sum_{\partidxi=1}^{N} q^{(\partidxi)} r_\partidxi^n Y_n^{-m}(\theta_\partidxi, \phi_\partidxi). 
\end{equation*}
This is converted into the local expansion $\Psi_{1,1}$ with expansion coefficients $H_n^m=\mathcal{R}\left(K_n^m\right)$ where $\mathcal{R}$ is the operator introduced in Section \ref{sec:boundary_conditions}. More specifically, the local expansion of the potential induced by the periodic images in $V^{(\lr)}$ is (c.f. Eq. \eqref{eqn:multipole_local_expansion})
\begin{equation}
    \varphi(r, \theta, \phi) = \sum_{n=0}^p \sum_{m=-n}^n H_n^m r^n Y_n^m(\theta, \phi).\label{eqn:local_far_field_expansion}
\end{equation}
Hence the energy of $N$ charges interacting with a far-field $\varphi$ which is expressed as the local multipole expansion in Eq. \eqref{eqn:local_far_field_expansion} is obtained by evaluating Eq. \eqref{eqn:local_far_field_expansion} at the particle positions $\vec{r}^{(\partidxi)}=(r_\partidxi,\theta_\partidxi,\phi_\partidxi)$, multiplying by $q^{(\partidxi)}$ and summing over all particles.  
\begin{align}
  U^{(\lr)} &= \sum_{i=1}^{N} q^{(\partidxi)} \sum_{n=0}^p \sum_{m=-n}^n H_n^m r_\partidxi^n Y_n^m(\theta_\partidxi, \phi_\partidxi)\label{eqn:local_H_sum}\\
    &= \sum_{n=0}^p \sum_{m=-n}^n E_n^m H_n^m,\quad
    \text{where}\quad E_n^m = \sum_{\partidxi=1}^{N} q^{(\partidxi)} r_\partidxi^n Y_n^m(\theta_\partidxi, \phi_\partidxi).\label{eqn:energy_as_dot}
\end{align}
The double sum in Eq. \eqref{eqn:energy_as_dot} is readily computed as a dot product in $(p+1)^2$ dimensions at a cost of $\mathcal{O}(p^2)$. Now consider a proposed move of a particle with charge $q$ from $\vec{r}$ to $\vec{r}'$. To evaluate the far-field energy of the modified charge distribution we need to do two things: first, we need to add $q{r'}^nY_n^m(\theta',\phi')$ to the expression for $E_n^m$ in Eq. \eqref{eqn:energy_as_dot} and simultaneously subtract $qr^n Y_n^m(\theta,\phi)$ since the position of the particle has changed in the sum in Eq. \eqref{eqn:local_H_sum}. Secondly, the far-field multipole coefficients $K_n^m$ and hence the local expansion coefficients $H_n^m$ in Eq. \eqref{eqn:local_far_field_expansion} change, since we assume that this far-field is generated by the periodic copies of the cell in $V^{(\infty)}$. This can be accounted for by adding a correction to the multipole expansion $K_n^m$ which describes a monopole of charge $+q$ at $\vec{r}'$ and a monopole of charge $-q$ at the old position $\vec{r}$. Those two modifications are achieved by setting
\begin{equation}
    \begin{aligned}
        \bar{E}_n^m &= E_n^m +  q\left({r'}^n Y_n^m(\theta', \phi') 
        -  r^n Y_n^m(\theta, \phi)\right)
    \end{aligned}
    \label{eqn:E_n_m_bar}
\end{equation}
and
\begin{equation}
    \begin{aligned}
        \bar{K}_n^m &= K_n^m 
        +  q\left({r'}^n Y_n^{-m} (\theta', \phi') 
        -  qr^n Y_n^{-m} (\theta, \phi)\right).
    \end{aligned}
    \label{eqn:K_n_m_bar}
\end{equation}
This gives the modified local expansion coefficients
\begin{equation*}
    \begin{aligned}
        \bar{H}_n^m &= \mathcal{R}\left(\bar{K}_n^m\right).
    \end{aligned}
\end{equation*}
The new far-field contribution to the system energy for the proposed move $\vec{r}'\leftarrow \vec{r}$ is
\begin{equation*}
    U^{(\lr)}_{\vec{r}'} = \sum_{n=0}^\infty \sum_{m=-n}^n \bar{E}_n^m \bar{H}_n^m,
\end{equation*}
\pparagraph{Total change in energy} 
Combining the near-field and far-field components the change in system energy of a proposed move $\vec{r}'\leftarrow \vec{r}$ is given by
\begin{equation*}
    \begin{aligned}
\Delta U_{\vec{r}'\leftarrow \vec{r}} &= \Delta U^{(\sr)}_{\vec{r}'\leftarrow \vec{r}} +  U^{(\lr)}_{\vec{r}'} - U^{(\lr)}_{\vec{r}}.
    \end{aligned}
\end{equation*}
The near-field component retains the $\mathcal{O}(p^2)$ computational complexity of the free-space case as the additional correction terms have a constant cost per proposal.
For the far-field energy evaluation, the creation of $\bar{E}$ and $\bar{H}$ involves the computation of $4p^2$ coefficients with complexity $\mathcal{O}(p^2)$.
The multipole-to-local operator $\mathcal{R}$ can be applied as a matrix-vector product with computational complexity $\mathcal{O}(p^4)$.
Finally, evaluation of the far-field local expansion exhibits a $\mathcal{O}(p^2)$ computational complexity.
Hence the overall cost of proposing a move is 
\begin{equation}
  \text{Cost}_{\text{propose}}^{(\text{BC})}=\mathcal{O}(p^4),\label{eqn:cost_propose_BC}
\end{equation}
independent of the number of charges. This should be compared to the $\mathcal{O}(p^2)$ complexity for free-space boundary conditions given in Eq. \eqref{eqn:cost_propose}.
\pparagraph{Accept move}
For an accepted move $\vec{r}'\leftarrow \vec{r}$ the local expansions on the finest level $\Psi_{L,\cellidxj}$ are updated as in the free-space case. This operation has a computational complexity of $\mathcal{O}(p^2 N)$.
To account for the far-field component, we store the quantities $E_n^m$ and $H_n^m$. Whenever a move $\vec{r}'\leftarrow \vec{r}$ is accepted, the values are updated according to Eqs. \eqref{eqn:E_n_m_bar} and \eqref{eqn:K_n_m_bar}, i.e.
\begin{xalignat*}{2}
    E_n^m &\leftarrow \bar{E}_n^m,&
    K_n^m &\leftarrow  \bar{K}_n^m.
\end{xalignat*}
This update requires the computation of $4p^2$ expansion terms, and hence has an $\mathcal{O}(p^2)$ cost.
For an accept operation in the fully periodic case the dominant cost is the update of the local expansions $\Psi_{L,\cellidxj}$ required for the near-field energy computation. We conclude that the total cost for accepting a move is
\begin{equation}
  \text{Cost}_{\text{accept}}^{(\text{BC})} = \mathcal{O}(p^2N).
\label{eqn:cost_accept_BC}
\end{equation}
which has the same computational complexity as the accept-step for free-space boundary conditions\footnote{To keep the interface in the code general and allow transitions to states which have not been previously proposed, the change in energy for the new state will be re-computed when accepting a move. This adds an additional $\mathcal{O}(p^4)$ cost, which can be safely neglected as long as $p^2\ll N$.} in Eq. \eqref{eqn:cost_accept}.

We conclude that the total complexity of the algorithm is $\mathcal{O}(p^4N)$ per KMC step even for non-trivial boundary conditions. The estimates in Eqs. \eqref{eqn:cost_propose_BC} and \eqref{eqn:cost_accept_BC} are confirmed numerically in Fig. \ref{fig:order_accept_propose} below.
\section{Related work}\label{sec:literature_review}
To highlight the impact of the novel FMM-based algorithm presented in this paper, we review existing approaches for including electrostatic interactions in KMC simulations.

As argued in \cite{Casalegno2010}, most recent KMC studies truncate or neglect long range electrostatic interactions. The few exceptions reported in the literature typically include some results obtained with the Ewald method, which serves as a computationally expensive reference implementation to quantify systematic errors introduced by those approximations.
For example, the authors of \cite{Hermet2013} do not include the effect of electrostatic interactions in most of their results for a KMC study in doped perovskites. This leads to a change in the protonic diffusion coefficient by $14\%$, compared to the ``correct'' result obtained with Ewald summation. A systematic comparison of photovoltaic simulations with truncated potentials and the Ewald method is also presented in \cite{Casalegno2010}. The authors find that introducing a cutoff potential underestimates the device performance and overestimates average charge carrier densities. The Ewald summation is optimised by precomputing the mutual potential between all pairs of charges. This reduces the cost of one proposal to $\mathcal{O}(N)$. On the structured lattice which is used in \cite{Casalegno2010} it requires the storage of $\mathcal{O}(N)$ terms, but for large systems with an irregular arrangement of the hopping sites the $\mathcal{O}(N^2)$ memory requirements would make the method infeasible. In \cite{Walls1999} perovskite crystal growth is modelled with a KMC method which uses Ewald summation for long range electrostatic interactions. However, the setup is very different to the problem considered here, since the system is described by a series of growing stacks on a 2d surface.

Electrostatic interactions can also be included by mapping the charge distribution to a grid and solving the Poisson equation. Since a solve of this three-dimensional partial differential equation is expensive, a lower dimensional approximation is used in some cases to make the computation feasible. For example, in \cite{vanderHolst2011,Kordt2015} the one dimensional Poisson equation is solved for a layer averaged charge density, while including the electrostatic field of nearby charges and periodic mirror images exactly. Naturally, this approach introduces uncontrollable systematic errors.

As discussed in \cite{Li2017}, including electrostatics by solving the three dimensional Poisson equation naively requires an expensive re-calculation of the potential for every potential hop since the charge has moved and its contribution can not be included in the current potential. To address this issue, the self-interaction error in the naive approach can be suppressed or removed by adding and subtracting the field of a single charge. While efficient methods (such as multigrid \cite{Trottenberg2000}) exist for solving the Poisson equation in $\mathcal{O}(n)$ time on a grid with $n$ cells (and it is reasonable to assume that the number of grid cells is at the same order as the number of lattice sites), discretising the Poisson equation is likely to lead to uncontrollable errors due to the peaked distribution of point particles, which can not be represented accurately on a grid. A massively parallel GPU implementation of a KMC method is described in \cite{vanderKaap2016}. While the authors still employ a cutoff for the electrostatic interactions, after each KMC step the local potential is updated by adding a dipole correction term, instead of recalculating the value of the field.
\pparagraph{Existing FMM implementations and KMC libraries}
Not surprisingly, there is a plethora of existing FMM implementations for the standard method written down in Algorithms \ref{alg:fmm} and \ref{alg:fmm_evaluation}. Some of those, such as ScalFMM \cite{Blanchard2015} and ExaFMM \mbox{\cite{Yokota2012}} are specifically designed for performance and massively parallel scalability; ExaFMM also targets GPU systems \cite{Yokota2013}. Similarly, a parallel FMM implementation for heterogeneous systems is described in \cite{Lashuk2009}. Other actively developed parallel libraries are FMMlib3d \cite{Gimbutas2015}, RECFMM \cite{Zhang2016} and DashMM \cite{DeBuhr2016}. Often those libraries can treat more general potentials, for example, ScalFMM is kernel-free and allows the user to implement their own interaction potentials. Since ScalFMM has been highly optimised for modern multicore processors, we quantify the absolute performance of our standard FMM implementation by comparing it to ScalFMM below. The PVFMM code \cite{Malhotra2015} includes several improvements, for example an $\mathcal{O}(p^3\log(p))$ FFT-based multipole-to-local translation and performance optisation through threading and vectorisation.
In FMM implementations the complexity of the multipole to local translation is of particular interest. In \cite{Greengard1997} Greengard et al.\ propose that, through a particular choice of the number of charges per cell on the finest level, all required multipole to local translations can be applied with a total cost that is $\mathcal{O}(N p^2)$.
For a comparison of different established FMM codes see also \cite{Yokota2015}. However, as far as we are aware, none of the above FMM libraries have been used in KMC simulations. Conversely, existing KMC libraries such as DL\_AKMC \cite{Gunn2014}, SPPARKS \cite{Plimpton2009} and KMCLib \cite{Leetmaa2014} do not include support for long range electrostatic interactions or rely on one of the approximations described above (see e.g. the study in \cite{Liu2017}, which uses the commercial Bumblebee library \cite{Bumblebee}).
\section{Implementation and user interface}\label{sec:implementation}
Algorithmically optimal methods such as the FMM-KMC scheme introduced in Section \ref{sec:FMMforKMC} have to be implemented efficiently on massively parallel hardware. With the recent diversification of the hardware landscape, it is equally important that the code has a simple, intuitive user interface and runs on different chip architectures, such as manycore CPUs and multithreaded GPUs. To ensure the reproducibility of our results and allow others to benefit from the library, we now describe the design of the code in some detail.
\subsection{Performance portable framework}
The FMM-KMC algorithms introduced in this paper were implemented on top of the performance portable framework described in \cite{Saunders2018}, which we refer to as ``PPMD'' in the following. The PPMD code is freely available at\vspace{-1ex}\begin{center}\url{https://github.com/ppmd/ppmd}.\end{center}\vspace{-1ex}
Our FMM-KMC implementation is provided in the \texttt{coulomb\_kmc} Python package and can be downloaded from\vspace{-1ex}\begin{center}\url{https://github.com/ppmd/coulomb_kmc}.\end{center}\vspace{-1ex}
A recent snapshot of the code, which can be used to reproduce the results in the paper, is also provided as \cite{fmmpaper_zenodo}.

The overall design principle of PPMD is to provide a high-level Python user interface which is flexible enough to express fundamental looping mechanisms for interacting particles, while automatically generating highly efficient code on different hardware platforms. 
As discussed in \secapp\ref{sec:KMC_bookkeeping}, using this Python interface also allows the easy and efficient implementation of the user-specific handling of the KMC data structures, such as masking forbidden moves based on the problem-dependent matrix of hopping sites.

Efficiency is achieved by using code generation techniques; once the user has expressed the fundamental interaction kernel as a short snippet of C-code, dedicated looping code which executes the entire particle- or particle-pair loop over all particles is generated. Depending in the platform, this might be realised with MPI, OpenMP or with CUDA on GPUs. High-level operations, such as the main time stepping loop, are implemented in Python and orchestrate the calls to the computationally expensive particle- or particle-pair loops. As shown in \cite{Saunders2018}, the performance of the PPMD code is on a par with monolithic MD codes written in C and Fortran, such as LAMMPS \cite{Plimpton1995} and DL\_POLY \cite{Smith1996,Todorov2006}.

While the main application of PPMD is molecular dynamics, the PPMD interface and data structures are abstract enough to also implement KMC algorithms which are the topic of this paper. As described in \cite{Saunders2017a}, PPMD also includes a library for calculating electrostatic interactions via Ewald summation; this is very useful for testing the FMM algorithms developed in this paper.
\pparagraph{Fundamental data structures and interfaces}
The most important data structure in PPMD is the \texttt{ParticleDat} class. This is a distributed storage space for data associated with individual particles in the simulation. Example properties which can be stored as \texttt{ParticleDat}s are the positions, velocities and charges of all particles in the system. However, in principle any property, such as the number of local neighbours of a particle, could be stored in a \texttt{ParticleDat}. Under the hood, \texttt{ParticleDat}s are realised as wrappers to numpy arrays. In addition, global properties shared by all particles, such as the total energy of the system, can be stored in \texttt{GlobalArray} or \texttt{ScalarArray} objects, where the latter is read-only when accessed from inside a kernel.

To manipulate data, the user writes a short kernel in C which is executed over all particles or all pairs of particles. A particle-pair loop is specified by this C-kernel and a list of all \texttt{ParticleDat}s, \texttt{ScalarArray}s and \texttt{GlobalArray}s which are modified by the kernel. Access descriptors specify whether the data is read or written. Based on this specification of the particle-loop, the code generation system automatically generates efficient code for executing the kernel over all pairs of particles. Depending on the access descriptors, it also performs suitable communication calls to synchronise parallel data access and avoids write conflicts in threaded implementations.

To illustrate the idea, we show a short Python code snippet for naively calculating the total potential energy $U$ for particles interacting via a Coulomb potential
\begin{equation}
  U = \frac{1}{2}\sum_{\substack{\text{all pairs}\\(\partidxi,\partidxj)}} \frac{q^{(\partidxi)} q^{(\partidxj)}}{\left|\vec{r}^{(\partidxi)}-\vec{r}^{(\partidxj)}\right|}\label{eqn:naive_coulomb}
\end{equation}
in code listing \ref{lst:naive_coulomb}. Note that the key operation
\begin{equation*}
U \mapsfrom U + 0.5\cdot q^{(\partidxi)} q^{(\partidxj)}/\left|\vec{r}^{(\partidxi)}-\vec{r}^{(\partidxj)}\right|
\end{equation*}
which is executed for all particle pairs is encoded in the C-code stored in the string \texttt{kernel\_code}. Positions and charges are stored in the \texttt{ParticleDat}s \texttt{r} and \texttt{q}, while the total energy is stored in the \texttt{GlobalArray} \texttt{U}. Interested readers are referred to \cite{Saunders2018} for more details on PPMD.
\begin{figure}
\begin{center}
\begin{minipage}{1.0\linewidth}
\begin{lstlisting}[language={[ppmd]{python}}, label=lst:naive_coulomb,caption={Python code for calculating the sum in Eq. (\ref{eqn:naive_coulomb}) over all particle pairs.}]
# number of particles
npart=1000

# Define Particle Dats
r = ParticleDat(npart=npart, ncomp=3, dtype=c_double)
q = ParticleDat(ncomp=1, npart=npart, dtype=c_double)
U = GlobalArray(ncomp=1, initial_value=0.0, dtype=c_double)

kernel_code='''
  double dr_sq = 0.0;
  for (int k=0;k<3;++k) {
    double dr = r.i[k]-r.j[k];
    dr_sq += dr*dr;
  }
  U += 0.5*q.i[0]*q.j[0]/sqrt(dr_sq);
'''

# Define kernel
kernel = Kernel('naive_coulomb',
                kernel_code)
  
# Define and execute pair loop
pair_loop = PairLoop(kernel=kernel,
                     {'r':r(access.READ), 'q':q(access.READ), 'U':U(access.INC)})
pair_loop.execute()
\end{lstlisting}
\end{minipage}
\end{center}
\end{figure}
\subsection{FMM-KMC user interface}
To implement the FMM algorithm for KMC simulations described in Section \ref{sec:FMMforKMC}, we introduce a new \texttt{KMCFMM} Python class in the \texttt{coulomb\_kmc} Python package.
This class provides three key operations:
\begin{enumerate}
\item \textbf{Initialisation} of the FMM fields and calculation of the system's full electrostatic energy at the start of the simulation
\item Evaluation of the electrostatic energy difference $\Delta U_{\vec{r}'\leftarrow\vec{r}}$ for all \textbf{proposals} $\vec{r}'\leftarrow\vec{r}$
\item \textbf{Acceptance} of a move selected by the user
\end{enumerate}
Note that the rest of the KMC algorithm, such as the selection of a move for acceptance based on the propensities and working out the set of allowed potential moves is still the responsibility of the user. Section \ref{sec:KMC_bookkeeping} discusses an example of how the latter can be realised efficiently with the PPMD data structures and parallel loops. Relegating high-level control over the algorithm to the user also allows the easy extension of the basic KMC method to more advanced techniques, such as multilevel methods, or the inclusion of post-processing steps to extract physically meaningful information.

The constructor of the \texttt{KMCFMM} class is passed the initial positions and charges of all particles, stored in \texttt{ParticleDat} objects. It allows the user to choose the parameters of the FMM solver, such as number of FMM levels and expansion terms. The \texttt{initialise()} method computes the system's electrostatic energy for the initial positions of all charges based on the standard FMM algorithm described in Algorithms \ref{alg:fmm} and \ref{alg:fmm_evaluation}. The method also creates and initialises all data structures required for the subsequent proposed and accepted moves.
\subsubsection{Proposing moves}\label{sec:proposing_moves}

The \texttt{KMCFMM} class provides two interfaces for calculating the change in system energy of proposed moves. The simple \texttt{propose()} method assumes that for a particle with index $\partidxi\in\{0,\dots,N-1\}$ located at position $\vec{r}^{(\partidxi)}$ and carrying charge $q_\partidxi$, there is a finite set of $n_\partidxi$ potential new positions $ \mathcal{R}^{(\partidxi)} = \{{\vec{r}'}^{(\partidxi)}_1, {\vec{r}'}^{(\partidxi)}_2, \dotsc,{\vec{r}'}^{(\partidxi)}_{n_i} \}$ which this particle could move to. The potential moves of all particles are stored as a list of tuples of the form $\left(\partidxi, \mathcal{R}^{(\partidxi)} \right)$ for $\partidxi\in\{\partidxi_1,\partidxi_2,\dots\}\subseteq \{0,\dots,N-1\}$, which are passed to the \texttt{propose()} method in the form
\begin{equation}
    \begin{aligned}
        \mathcal{P} = \left( \left(\partidxi_1, \mathcal{R}^{(\partidxi_1)} \right),
          \left(\partidxi_2, \mathcal{R}^{(\partidxi_2)} \right), \dotsc \right).
    \end{aligned}\label{eqn:tuple_P}
\end{equation}
The change in total electrostatic energy for the move ${\vec{r}'}^{(\partidxi)}_k\leftarrow \vec{r}^{(\partidxi)}$ with $k\in\{1,\dots,n_\partidxi\}$ is denoted by $\Delta U_{{\vec{r}'}_k^{(\partidxi)}}$. The changes in energy for all potential moves of particle $\partidxi$ are collected in the list
\begin{equation}
\mathcal{U}^{(\partidxi)} = \{ \Delta U_{{\vec{r}'}^{(\partidxi)}_1}, \Delta U_{{\vec{r}'}^{(\partidxi)}_2}, \dotsc, \Delta U_{{\vec{r}'}^{(\partidxi)}_{n_\partidxi}}\}.
\label{eqn:setU}
\end{equation}
The \texttt{propose()} method returns a tuple of arrays of electrostatic energy differences $\left( \mathcal{U}^{(\partidxi_1)} , \mathcal{U}^{(\partidxi_2)}, \dotsc \right)$, where each $\mathcal{U}^{(\partidxi)}$ is of the form described in Eq. \eqref{eqn:setU}.

To encode three-dimensional vectors, the proposed new positions $\mathcal{R}^{(\partidxi)}$ are passed to the \texttt{propose()} method as a $n_\partidxi\times 3$ dimensional \texttt{numpy} array for each charge. The method returns a \texttt{numpy} array with the change in electrostatic energy for each proposed move. An example is shown in code listing \ref{lst:kmc_propose}. In this case particle 3 can move to two potential new positions, namely ${\vec{r}'}_1^{(3)}=(0.11,0.13,0.09)$ and ${\vec{r}'}_2^{(3)}=(0.90,0.10,0.08)$, whereas there is only one potential hop to the new position ${\vec{r}'}_1^{(7)}=(0.45,0.28,0.89)$ for particle 7.
For this setup the tuple $\mathcal{P}$ is given by
\begin{equation*}
\begin{aligned}
\mathcal{P} = \Big( &
\left(3,
\begin{bmatrix}
0.11 & 0.13 & 0.09\\
0.90 & 0.10 & 0.08
\end{bmatrix}
\right),
\left(
7,\begin{bmatrix}
0.45 & 0.28 & 0.89
\end{bmatrix}
\right)
\Big)
\end{aligned}
\end{equation*}
which should be compared to the Python code in Listing \ref{lst:kmc_propose}. After the call to \texttt{propose()}, the variable \texttt{U} will contain the corresponding energy differences as a tuple of numpy arrays in the format
\begin{equation*}
U = \left(
\left[\Delta U_{{\vec{r}'}_1^{(3)}},\Delta U_{{\vec{r}'}_2^{(3)}}\right],
\left[\Delta U_{{\vec{r}'}_1^{(7)}}\right]
\right).
\end{equation*}
\begin{figure}
\begin{minipage}{\linewidth}
    \begin{lstlisting}[language={[ppmd]{python}}, label=lst:kmc_propose,caption={Python code for creating a \texttt{KMCFMM} instance and proposing moves as described in Section \ref{sec:proposing_moves}.}]
import numpy as np

# Create KMCFMM instance
kmc_fmm = KMCFMM(positions=r, charges=q, r=4, l=12)

# Set up FMM data structure, calculate
# initial electrostatic energies
kmc_fmm.initialise()                 

# Tuple with proposed moves
P = (
      (3, np.array(((0.11, 0.13, 0.09), (0.90, 0.10, 0.08)))  # Particle 3
      ),
      (7, np.array(((0.45, 0.28, 0.89),))                     # Particle 7
      )
    )

# Calculate energy changes 
U = kmc_fmm.propose(P)
\end{lstlisting}
\end{minipage}
\end{figure}
While the interface for proposing moves described here is intuitive, it is not optimal in terms of efficiency. The improved, but more sophisticated, \texttt{propose\_with\_dats()} interface is described in \secapp\ref{sec:improved_proposal_interface}, where we also illustrate its use for a practically relevant example.
\subsubsection{Accepting moves}\label{sec:accepting_moves}
The \texttt{KMCFMM} instance provides a \texttt{accept} method that accepts a proposed move, updates the system energy and updates internal data structures.
To accept a move, the \texttt{accept()} is called with a tuple $\left(\partidxi,\vec{r}'\right)$ consisting of a charge index $\partidxi$ and a new position $\vec{r}'$. An example is given in code listing \ref{lst:kmc_accept}, which assumes that particle $\partidxi=6$ moves to the new position $\vec{r}'=(0.11,0.56,0.39)$.
\begin{figure}
\begin{minipage}{\linewidth}
\begin{lstlisting}[language={[ppmd]{python}}, label=lst:kmc_accept,caption={Python code for accepting a move as described in Section \ref{sec:accepting_moves}.}]
p = kmc_fmm.accept((6, np.array((0.11,0.56,0.39))))
\end{lstlisting}
\end{minipage}
\end{figure}
\subsection{Parallelisation and optimisation}\label{sec:parallelisation}
On distributed memory machines domain-de\-com\-pos\-it\-ion is used to parallelise the KMC-FMM algorithm described in Section \ref{sec:FMMforKMC}. For this the computational domain $\Omega$ is divided between MPI ranks such that each rank ``owns'' the charges in its local subdomain $\Omega_{\text{local}}$. Proposals for particles owned by different MPI ranks are handled concurrently. $\Omega_{\text{local}}$ is augmented by a halo region to obtain an extended local domain $\overline{\Omega}_{\text{local}}$. It is assumed that the particles are evenly distributed and hops are limited by some maximal distance which determines the size of this halo and hence $\overline{\Omega}_{\text{local}}$. Under those conditions, which seem reasonable for many physical systems, domain decomposition will result in good load balancing and -- as will be described in the following -- requires very little parallel communication. The results in Section \ref{sec:results_MPI} confirm the excellent parallel scaling on up to 128 nodes.

To store local- and multipole expansions $\Psi_{\ell,\cellidxi}$ and $\Phi_{\ell,\cellidxi}$ on all levels of the grid hierarchy, a distributed ``Octal Tree'' (OT) data structure is set up at the beginning of the simulation. As described in Section 5.3 of \cite{Saunders2018a}, data can be attached to the OT using different parallel access modes. This allows suitable halo-exchanges between neighbouring MPI ranks and ensures that all children of a particular coarse level cell are owned by the same rank during the upward- and downward pass of Algorithm \ref{alg:fmm}. The local expansions $\Psi_{L,\cellidxi}$ are calculated for all cells $\cellidxi$ on the finest level at the beginning of the simulation. The OT cells on the finest level are not necessarily aligned with the local subdomains $\Omega_{\text{local}}$. However, each MPI rank keeps copies of the local expansions $\Psi_{L,\cellidxi}$ for all fine level OT cells which cover its extended domain $\overline{\Omega}_{\text{local}}$. It also maintains copies of all particle positions and charges in those cells. 

When a potential move $\vec{r}'\leftarrow\vec{r}$ is proposed, calculating the change in energy $\Delta U_{\vec{r}'\leftarrow\vec{r}}$ in Eq. \eqref{eqn:proposed_energy} for free-space boundary conditions requires the evaluation of the local expansion on the finest level $\Psi_{L,\cellidxi}$ at the old and new positions, and knowledge of particle positions in neighbouring cells. This data is available in local copies, provided the halo is chosen large enough. Upon accepting a move, ownership is transferred if the moved particle crosses a subdomain boundary. The local expansions $\Psi_{L,\cellidxi}$ are updated for all cells $\cellidxi$ which are affected by this move. Note that neither evaluating the energy change for the proposals nor the update of $\Psi_{L,\cellidxi}$ at the end of a KMC step requires any parallel communication apart from sharing the details of the accepted move between all processors.

For non-trivial boundary conditions the \textit{near-field} change in electrostatic energy $\Delta U_{\vec{r}'\mapsfrom \vec{r}}^{(\sr)}$ given by Eq. \eqref{eqn:proposed_energy_sr} can be handled in the same way as just described for free-space boundary conditions, both when proposing and accepting a move. Calculating the change in the \textit{far-field} contribution $U^{(\lr)}$ given by Eq. \eqref{eqn:local_H_sum} requires an update to the expansion coefficients $H_n^m$ on the coarsest level and $E_n^m$ defined in Eq. \eqref{eqn:energy_as_dot}. Identical copies of both $H_n^m$ and $E_n^m$ are stored on all MPI ranks. When a move is accepted, they are updated locally on each MPI rank by removing the contribution from the particle at $\vec{r}$ and adding the contribution from the new position $\vec{r}'$.

We use OpenMP as a shared memory programming method to distribute the proposed moves on each MPI rank over available cores. This reduces load imbalances due to variations in the number of proposed moves per particle. Furthermore, in this hybrid MPI+OpenMP approach the volumes of the subdomains handled by individual MPI ranks are larger than in an MPI only execution. In addition to improved load-balancing, this increases the ratio of subdomain volume to halo region volume, which reduces the computational work of accepting a move as fewer expansions and particles must be updated.

All performance critical operations which manipulate multipole and local expansions are implemented as auto-generated C code. This allows fixing the number of expansion coefficients $p$ at compile-time, which enables important optimisations such as loop-unrolling and auto-vectorisation. In particular, evaluation of the spherical harmonics $Y_\ell^m(\theta,\phi)$ at a particular position $\vec{r}=(r,\theta,\phi)$ with recurrence relations (see e.g. \cite{Abramowitz1965,Press2007}) is carried out with a two-level loop nest. The bounds of the inner loop with index $k$ depend on the current iteration of the outer loop over $\ell\in 0,\dots,p$ which makes it virtually impossible to vectorise the code if the outer loop bound $p$ is only fixed at runtime. However, if $p$ is known at compile time, the loop can be unrolled to generate a long sequential list of updates, which contain the same algebraic operations and can are readily vectorised. In addition, when generating this code combinatorial factors which depend on the loop indices $k$ and $\ell$ can be pre-computed at compile time. This reduces the number of floating point operations and further improves performance.

Finally, note that any further book-keeping operations required in a KMC step, such as those described in Algorithm \ref{alg:kmc_ppmd}, are implemented as \texttt{ParticleLoop} and \texttt{PairLoop} constructs in PPMD. They are therefore automatically parallelised on any parallel architecture, as described in detail in \cite{Saunders2018a, Saunders2018}.
\section{Results}\label{sec:results}
The KMC simulations reported in this section were carried out on the CPU nodes of the ``Balena'' cluster at the University of Bath, with some additional weak- and strong- parallel scaling runs on the ARM-based ``Isambard'' supercomputer and the Intel-based CSD3 Peta4 computer. 
On Balena, Intel Ivy Bridge and Skylake nodes with 16 and 24 cores in total were used. On Isambard, one full Cavium ThunderX2 node contains 64 cores. Finally, on CSD3 Peta4 a full node contains 32 cores.
The code was run in mixed MPI+OpenMP mode, the exact node configuration and process layout can be found in Tab. \ref{tab:node_layout}. This setup was used for all runs, unless explicitly stated otherwise below. All autogenerated code in PPMD was compiled with version 17.1.132 of the Intel compiler on Balena, using the same version of IntelMPI for distributed memory parallelisation. On Isambard GCC version 8.2.0 was used together with CRAY MPICH 7.7.6. Finally, on CSD3 Peta4, Intel compiler and IntelMPI version 2019.3 was used.
Raw results and code snapshots are publicly available for reproducibility purposes in the archive at \cite{fmmpaper_zenodo}.

\begin{table*}
\begin{center}
\begin{tabular}{cp{0.5ex}cccccc}
\hline
machine & & chip & sockets & cores per  & cores per & MPI   & OpenMP threads \\
        & &      &         & socket & node   & ranks & per MPI rank\\
\hline\hline
\multirow{2}{*}{Balena} & \multirow{2}{*}{$\Big\{$} & \multicolumn{1}{l}{Intel Ivy Bridge E5-2650v2} & 2 & 8 & 16 & 4 & 4\\
 & & \multicolumn{1}{l}{Intel Skylake Gold 6126} & 2 & 12 & 24 & 4 & 6\\
Isambard & & \multicolumn{1}{l}{Cavium ThunderX2} & 2 & 32 & 64 & 8 & 8\\
CSD3 Peta4 & & \multicolumn{1}{l}{Intel Skylake Gold 6142} & 2 & 16 & 32 & 8 & 4\\
\hline
\end{tabular}
\end{center}
\caption{Node configuration and process layout used for performance measurements on the Balena and Isambard machines.}
\label{tab:node_layout}
\end{table*}
\subsection{Error analysis}
We first demonstrate that the errors in the FMM-KMC method can be systematically quantified and decrease as the number $p$ of multipole expansion coefficients increases. In Chapter 5 of \cite{Saunders2018a} we measured the error on the \textit{total} electrostatic system energy $U$ for the standard FMM algorithm and studied its dependency on $p$. In a KMC simulation, however, the quantity of interest is the \textit{change} $\Delta U$ in the system's electrostatic energy for each proposed move. As confirmed by the results in Tab. \ref{tab:error_diffs}, $\Delta U$ is typically several (two or more) orders of magnitude smaller than $U$, which is proportional to the problem size. This places tighter bounds on the allowed numerical errors, which have to be small compared to the energy change $\Delta U$ for individual proposed moves.

To quantify the relative error on $\Delta U$, we consider a system of constant density (0.01 charges per unit volume) and record the change in energy for $M=10^4$ proposed moves. As in \cite{Saunders2018a}, the charges are initially arranged in an almost regular cubic lattice with small random perturbations to create a setup which is representative of physically relevant systems. Each proposed move corresponds to an additional small arbitrary displacement from the initial positions. To explore a wide range of configurations, a new pseudo-random system was generated every $100$ proposals.

For each proposed move $m$ we calculate a highly accurate estimate of the ``true'' energy difference $\Delta U_m^*$ by using our KMC-FMM algorithm with 26 expansion terms. The corresponding energy difference computed with $p<26$ expansion terms is denoted as $\Delta U_m$. Based on this we define the absolute relative error of move $m$ as
\begin{equation}
    \delta U_m = \frac{|\Delta U_m - \Delta U_m^*|}{|\Delta U_m^*|}
\label{eqn:diff_deltaUi}
\end{equation}
and use this quantity to assess the accuracy of the simulation.

\begin{figure}
\begin{center}
\includegraphics[width=0.5\linewidth]{\figdir/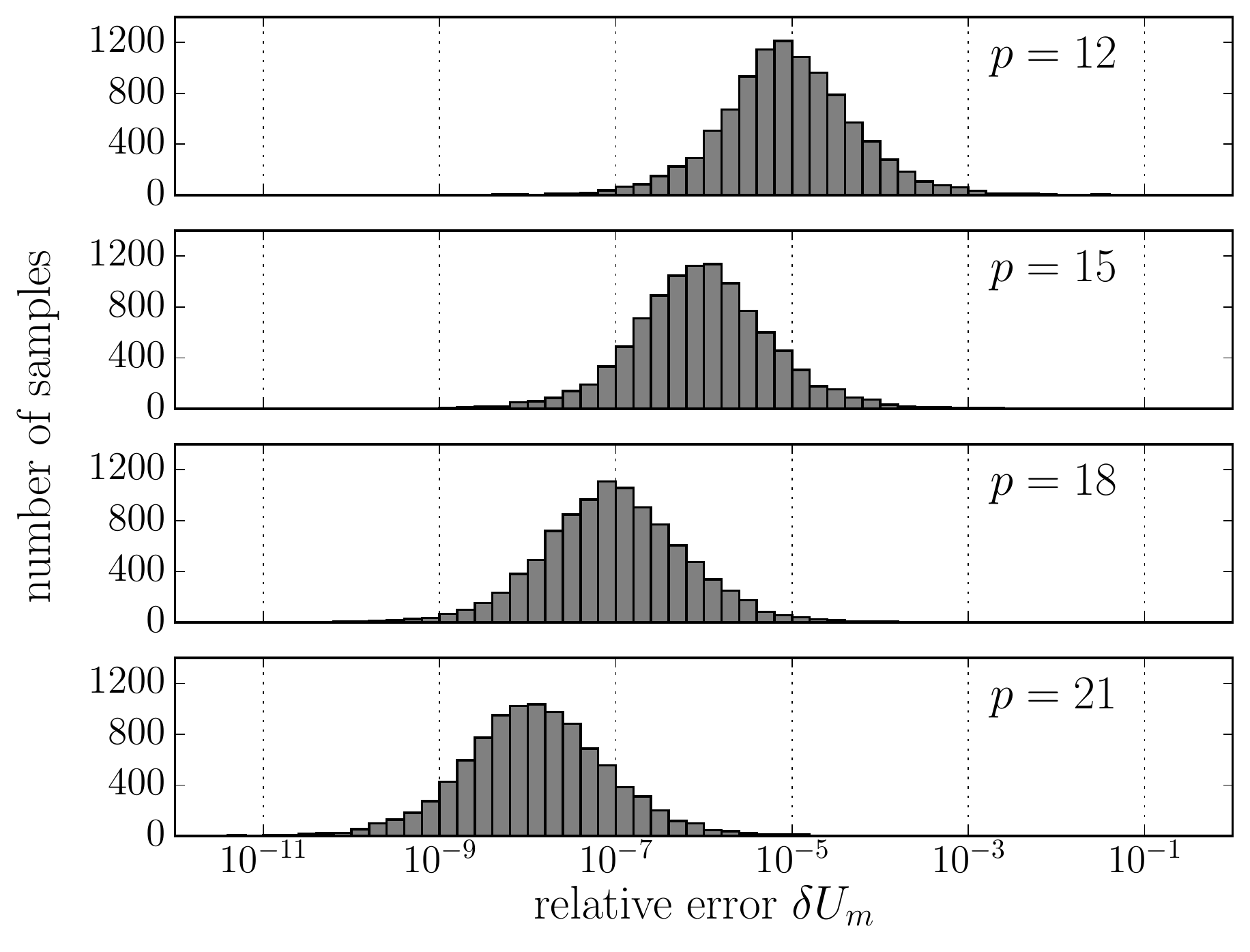}
\caption{Histogram of the relative error $\delta U_m$ defined in Eq. \eqref{eqn:diff_deltaUi} for different numbers of expansion terms $p$ and a system with $N=10^4$ charges.}
\label{fig:error_histogram}
\end{center}
\end{figure}

We initially keep the system size fixed at $N=10^4$ and vary the number of expansion terms $p$ from 12 to 21; the histograms in Fig. \ref{fig:error_histogram} show the number of samples with a given error for increasing $p$. Inspecting the distribution plotted there with logarithmic scale on the horizontal axis, it can be seen that the error on $\Delta U$ is reduced by around one order of magnitude as $p$ is incremented by 3.
To further quantify the size of the error, let
\begin{equation*}
  \langle A \rangle \equiv \frac{1}{M}\sum_{m=1}^{M} A_m
\end{equation*}
be the sample average over $M$ independent proposed moves for some quantity $A$.
We estimate the expected relative error and its variance as
\begin{xalignat}{2}
    \mathbb{E}(\delta U) &\approx \langle \delta U \rangle, &
    \text{Var}(\delta U) &= \mathbb{E}\left((\delta U-\mathbb{E}(\delta U))^2\right) \approx 
    \langle \delta U^2 \rangle - \langle \delta U \rangle^2
\label{eqn:diff_var_errvar}
\end{xalignat}
Corresponding estimates for the average absolute energy difference $|\Delta U^*|$ and the average absolute system energy $|U|$ are\footnote{Since the system is initialised at random, there is an equal probability of $U$ to be positive or negative, and it is therefore natural to consider its absolute value.}
\begin{xalignat*}{2}
    \mathbb{E}(|\Delta U^*|) &\approx \langle|\Delta U^*|\rangle, &
    \mathbb{E}(|U|) &\approx \langle|U|\rangle.
\end{xalignat*}
The dependency of $\langle \delta U\rangle $ on the number of expansion terms is shown in Tab.\ \ref{tab:error_p_diffs}, where we also we give the estimated variance, $\langle \delta U^2\rangle - \langle \delta U\rangle^2$ and other relevant quantities for $N=10^4$ charges and $p \in \{ 12, 15, 18, 21\}$ expansion terms. 
\begin{table*}[]
\begin{center}
\begin{tabular}{ccccc}
\hline
$p$ & $\langle \delta U\rangle$ & $\langle \delta U^2\rangle-\langle \delta U \rangle^2$ & $\langle|\Delta U^*|\rangle$ & $\langle|U|\rangle$ \\
\hline
\hline
$12$ & $8.93\cdot 10^{-5}$ & $2.30\cdot 10^{-6}$  & \multirow{4}{*}{\begin{minipage}{1.5cm}\begin{center}$\vert\vert$\\[0.5ex]$1.51\cdot 10^{-1}$\\[0.5ex]$\vert\vert$\end{center}\end{minipage}} & \multirow{4}{*}{\begin{minipage}{1.5cm}\begin{center}$\vert\vert$\\[0.5ex]$9.13\cdot 10^{2}$\\[0.5ex]$\vert\vert$\end{center}\end{minipage}}\\
$15$ & $8.66\cdot 10^{-6}$ & $1.14\cdot 10^{-8}$ \\
$18$ & $1.24\cdot 10^{-6}$ & $6.63\cdot 10^{-10}$ \\
$21$ & $1.81\cdot 10^{-7}$ & $8.82\cdot 10^{-12}$ \\
\hline
\end{tabular}
    \caption{Sample average and variance of the relative error $\delta U$ on the energy difference $\Delta U$, as defined in Eqs. \eqref{eqn:diff_deltaUi} and \eqref{eqn:diff_var_errvar}. Results are shown for a varying number of expansion terms $p$ and fixed $N=10^4$. The last two columns also give the sample average of the ``true'' energy change per proposed move $\Delta U^*$ and the total electrostatic energy $U$ of the system.}
\label{tab:error_p_diffs}
\end{center}
\end{table*}
We repeated the same experiment for fixed $p=12$ and varying problem sizes $N$, the results are shown in Tab. \ref{tab:error_diffs}.
\begin{table*}[]
\begin{center}
\begin{tabular}{ccccc}
\hline
    $N$ & $\langle \delta U\rangle$ & $\langle \delta U^2\rangle-\langle \delta U \rangle^2$ & $\langle|\Delta U^*|\rangle$ & $\langle|U|\rangle$ \\
\hline
\hline
$10^3$ & $1.58\cdot 10^{-4}$ & $7.96\cdot 10^{-5}$  & $1.41\cdot 10^{-1}$ & $8.73\cdot 10^{1}$ \\
$10^4$ & $8.93\cdot 10^{-5}$ & $2.30\cdot 10^{-6}$  & $1.51\cdot 10^{-1}$ & $9.13\cdot 10^{2}$ \\
$10^5$ & $8.07\cdot 10^{-5}$ & $4.36\cdot 10^{-7}$  & $1.13\cdot 10^{-1}$ & $1.81\cdot 10^{4}$ \\
\hline
\end{tabular}
    \caption{Sample average and variance of the relative error $\delta U$ on the energy difference $\Delta U$, as defined in Eqs. \eqref{eqn:diff_deltaUi} and \eqref{eqn:diff_var_errvar}. Results are shown for fixed $p=12$ and different problem sizes $N$. The last two columns also give the sample average of the ``true'' energy change per proposed move $\Delta U^*$ and the total electrostatic energy $U$ of the system.}
\label{tab:error_diffs}
\end{center}
\end{table*}
As confirmed by those tables, on average the relative error $\delta U$ defined in Eq. \eqref{eqn:diff_deltaUi} is about three orders of magnitude smaller than the average change in energy $\Delta U$ itself. The total electrostatic energy of the system grows in proportion with the number of charges and is significantly larger than $\Delta U$.
\subsection{Computational complexity}
\begin{figure}
\begin{center}
\begin{minipage}{0.45\linewidth}
\begin{center}
    \includegraphics[width=1.00\linewidth]{\figdir/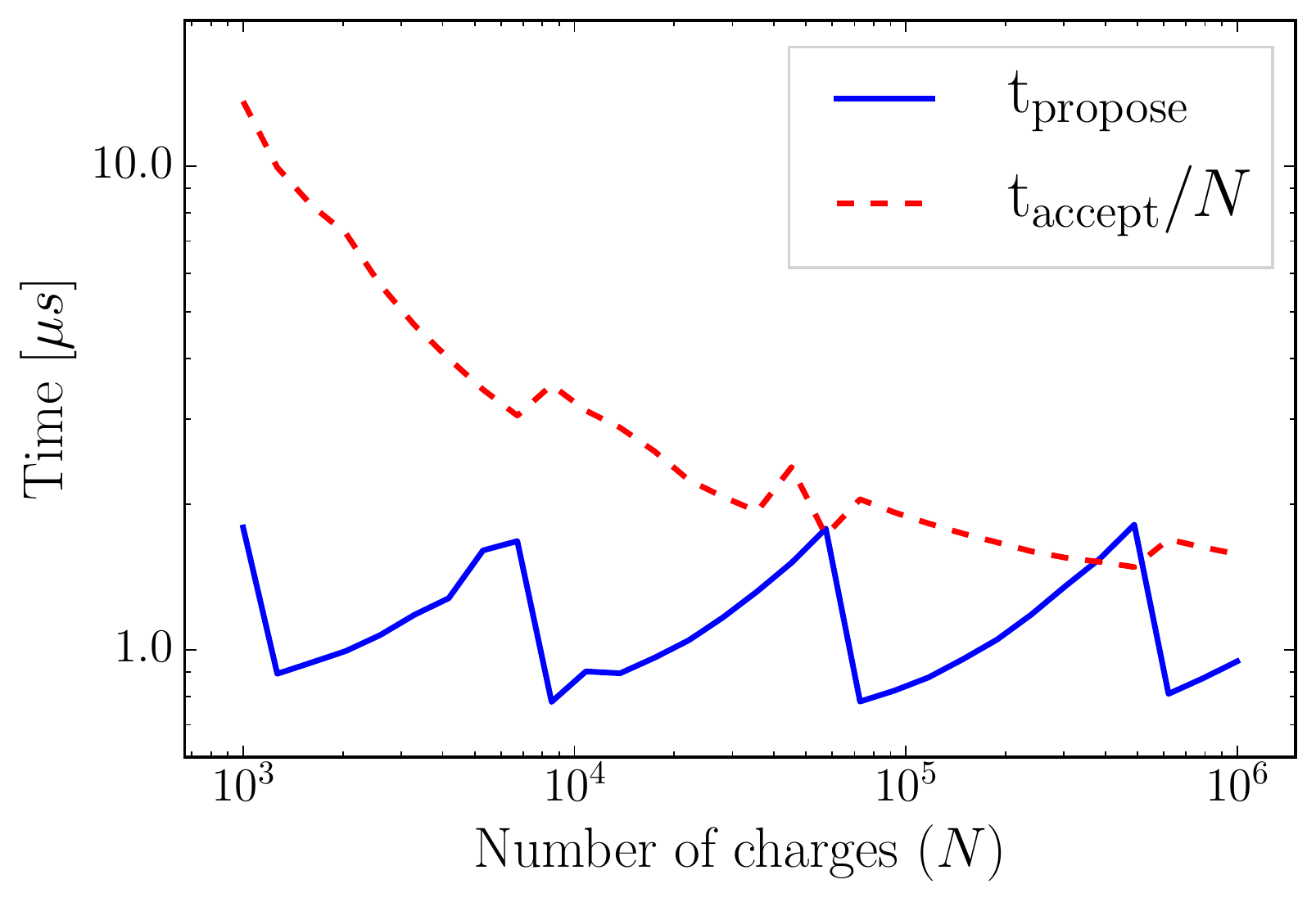}
    \caption{Time per proposed move $t_{\text{propose}}$ and time for accepting a proposal per particle $t_{\text{accept}}/N$ as a function of the number of charges. All results were obtained on a single Ivy Bridge node. The fluctuations in $t_{\text{accept}}/N$ are discussed in the main text.}
    \label{fig:propose_accept}
\end{center}
\end{minipage}\hfill
\begin{minipage}{0.45\linewidth}
\begin{center}
    \includegraphics[width=1.00\linewidth]{\figdir/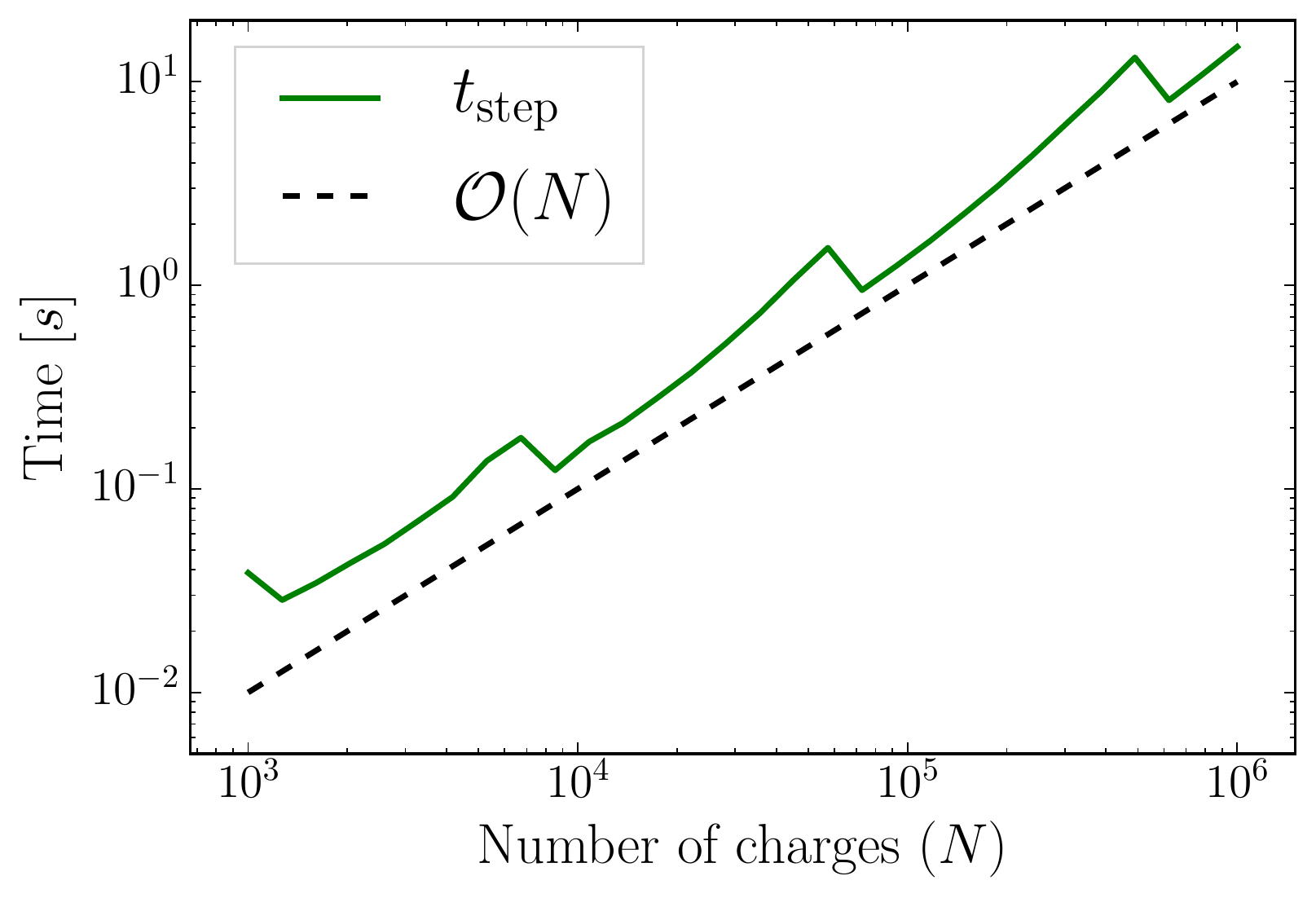}
    \caption{Total time per KMC $t_{\text{step}}$ estimated using Eq. \eqref{eqn:t_KMCstep} and the numerical values for $t_{\text{propose}}$ and $t_{\text{accept}}$ in Fig. \ref{fig:propose_accept}, which were obtained on a single Ivy Bridge node. The straight dashed line shows ideal linear scaling with the number of charges $N$.}
    \label{fig:time_kmc_step}
\end{center}
\end{minipage}
\end{center}
\end{figure}

Next we confirm that the runtime of the method grows in direct proportion to the number of charges $N$. We consider a system with periodic boundary conditions. Recall that theoretically proposing a single move carries a cost of $\mathcal{O}(p^4)$ and accepting a move costs $\mathcal{O}(Np^2)$, resulting in a linear growth of the computational complexity per KMC step. This is demonstrated by varying the number of charges in a system (for a fixed number $p=12$ of expansion terms) and plotting the time $t_{\text{propose}}$ for an individual proposal and for accepting a move (normalised to the number of charges) $t_{\text{accept}}/N$ in Fig. \ref{fig:propose_accept}. As this figure shows, both times are in the range of a few $\mu \text{s}$ when running the code on 16 cores of an Intel Ivy Bridge node.

The sawtooth nature of the plot is an artifact of the varying number of FMM levels, which depends on the problem size $N$ as $L = \min(3, \lfloor\log_8(N/2)\rfloor)$. The sharp drops on the right edge of the ``teeth'' correspond to an increase of $L$ by one, whereas all points on the left, shallow side of the ``teeth'' were obtained with the same value of $L$, which becomes less optimal as the problem size grows.
Based on those numbers, we estimate the total time per KMC step as
\begin{equation}
  t_{\text{step}} = N\cdot \overline{n}_{\text{propose}}\cdot t_{\text{propose}} + t_{\text{accept}}.\label{eqn:t_KMCstep}
\end{equation}
where $\overline{n}_{\text{propose}}=14$ is the estimated average number of proposed moves per charge and per KMC step. This value is motivated by the setup in Section \ref{sec:results_MPI}, and of the same order of magnitude as observed values for the $\alpha$-NPD test case in Section \ref{sec:results_alphaNPD}. The time per KMC step $t_{\text{step}}$ is plotted Fig. \ref{fig:time_kmc_step}, which confirms that our implementation indeed achieves the expected $\mathcal{O}(N)$ computational complexity.

The results imply that it is possible to simulate a system with one million charges in about $10\text{s}$ per KMC step when running on a single 16 core Ivy Bridge node and limiting the relative error in the energy difference per proposed move to $\sim 10^{-3}$.

\begin{figure}
\begin{center}
    \includegraphics[width=0.45\linewidth]{\figdir/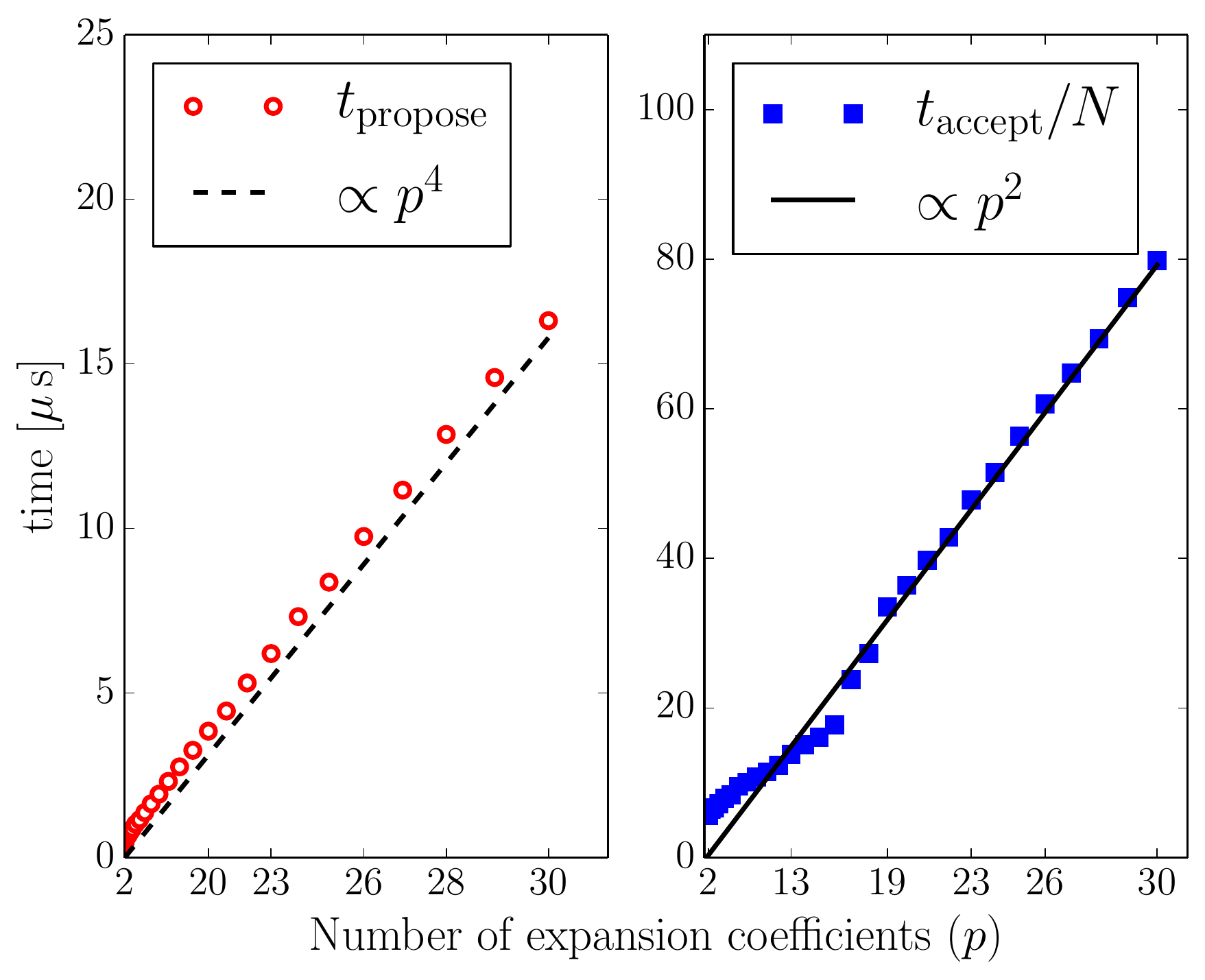}
    \caption{Time per proposal (left) and per acceptance step normalised to $N$ (right) for varying polynomial degrees $p$ and fixed $N=10^5$. The horizontal axes are scaled to highlight the quartic/quadratic dependency on $p$. All results are obtained on an Ivy Bridge node.}
    \label{fig:order_accept_propose}
\end{center}
\end{figure}

To demonstrate that the computational complexity depends polynomially on the number of expansion terms $p$, we repeated the above experiment but fixed the number of charges at $N=10^5$ while varying $p$ between 2 and 30.
The measured runtimes in Fig. \ref{fig:order_accept_propose} confirm that $t_{\text{accept}}/N$ depends quadratically on the number of expansion terms, whereas $t_{\text{propose}}$ is asymptotically proportional to the fourth power of $p$.
For the following numerical experiments $p=12$ expansion terms were used.
\subsection{Distributed Memory Parallelism}\label{sec:results_MPI}
While the previous section shows that it is possible to carry out KMC simulations with one million charged particles in a reasonable time on a single Ivy Bridge node, physically meaningful results can often only be extracted from much larger systems. Setups with $10^6-10^9$ particles allow the resolution of grain boundaries and ultimately move closer to the simulation of full photovoltaic devices and batteries. Modelling systems of these sizes in reasonable times requires distributed memory parallelism to utilise multiple compute nodes in a HPC facility.

As discussed in Section \ref{sec:parallelisation}, our implementation employs a hybrid MPI+OpenMP parallelisation strategy. To demonstrate the parallel scaling of a full KMC simulation we implemented the book-keeping algorithm in PPMD, employing the same techniques as in the example described in \secapp\ref{sec:KMC_bookkeeping}. Recall that this uses the \texttt{propose\_with\_dats()} interface of our FMM-KMC implementation for optimal efficiency. The energy differences $\Delta U_{\vec{r}'\leftarrow\vec{r}}$ of all proposed moves $\vec{r}'\leftarrow\vec{r}$ are used to calculate the associated propensities $\propto \exp\left( - \Delta U_{\vec{r}' \leftarrow \vec{r}} \right)$ by using a \texttt{ParticleLoop}. To choose a transition following steps 5-7 of the KMC Algorithm \ref{alg:KMC}, the partial sums $R_{\stateidxi\stateidxj}$ are computed on each MPI rank and combined across all processes using an \texttt{MPI\_Allgather} operation. Finally, the chosen proposed move is accepted by all MPI ranks.

To investigate how effectively our implementation scales across multiple compute nodes we performed both weak- and strong- scaling experiments. In the strong scaling experiment the problem size is kept fixed while the number of compute nodes is increased, resulting in a reduction of the total runtime. Since the local problem size decreases and hence the ratio between communication and computation usually gets worse, strong scaling is typically harder to achieve. In contrast, for the weak scaling experiment the problem size is increased in proportion to the number of nodes, in other words the \textit{local} problem size remains constant while the total \textit{global} problem size grows.

In both cases we consider a cubic lattice with a spacing of $h=1.1\text{\AA}$ in each direction such that one in 27 sites is occupied by a charged particle. We allow proposed moves to the 14 neighbouring lattice sites which are either a distance $h$ or $\sqrt{3}h$ away. In units of the lattice spacing the corresponding offset vectors are $(1,0,0)$, $(-1,0,0)$, $(0,1,0)$, \dots, $(1,1,1)$, $(-1,1,1)$ etc. and proportional to the lattice vectors of a fcc crystalline structure. Periodic boundary conditions are applied for the electrostatic field.

For the strong scaling experiments the global problem on a single node contains $10^6$ charges in total, corresponding to a cubic lattice with 300 points in each coordinate direction. On the largest node count (128 nodes on Isambard), this results in a local problem size with just under 7,800 charges per node. In the weak scaling experiment each node is responsible for $10^6$ charges on average. The largest problem which was simulated contains $1.28\cdot 10^8$ charges. 

Both scaling experiments were carried out on the hardware listed in Tab. \ref{tab:node_layout}, comparing fully populated nodes on the different machines.
For the strong scaling experiment the number of FMM levels was fixed at $L = 6$. For the weak scaling experiment the number of levels was chosen as $L = \lfloor \log_8(\gamma N) \rfloor$ where $\gamma$ is tuned for each machine and depends on the relative cost of the direct interactions (lines 5-7 in Algorithm \ref{alg:fmm}) and calculation of $\Psi_{L,\alpha}$ (Algorithm \ref{alg:fmm_evaluation} and line 4 in Algorithm \ref{alg:fmm}) on a particular hardware. On Isambard a value of $\gamma=3.4$ was found to be optimal, whereas $\gamma=4.7$ and $\gamma=3.3$ turned out to give the best results on the Ivy Bridge and Skylake nodes of Balena respectively. Finally, for CSD3 we determined $\gamma=4.19$ to be optimal.

\begin{figure}
\begin{center}
\begin{minipage}{0.576\linewidth}
\begin{center}
    \includegraphics[width=\linewidth]{\figdir/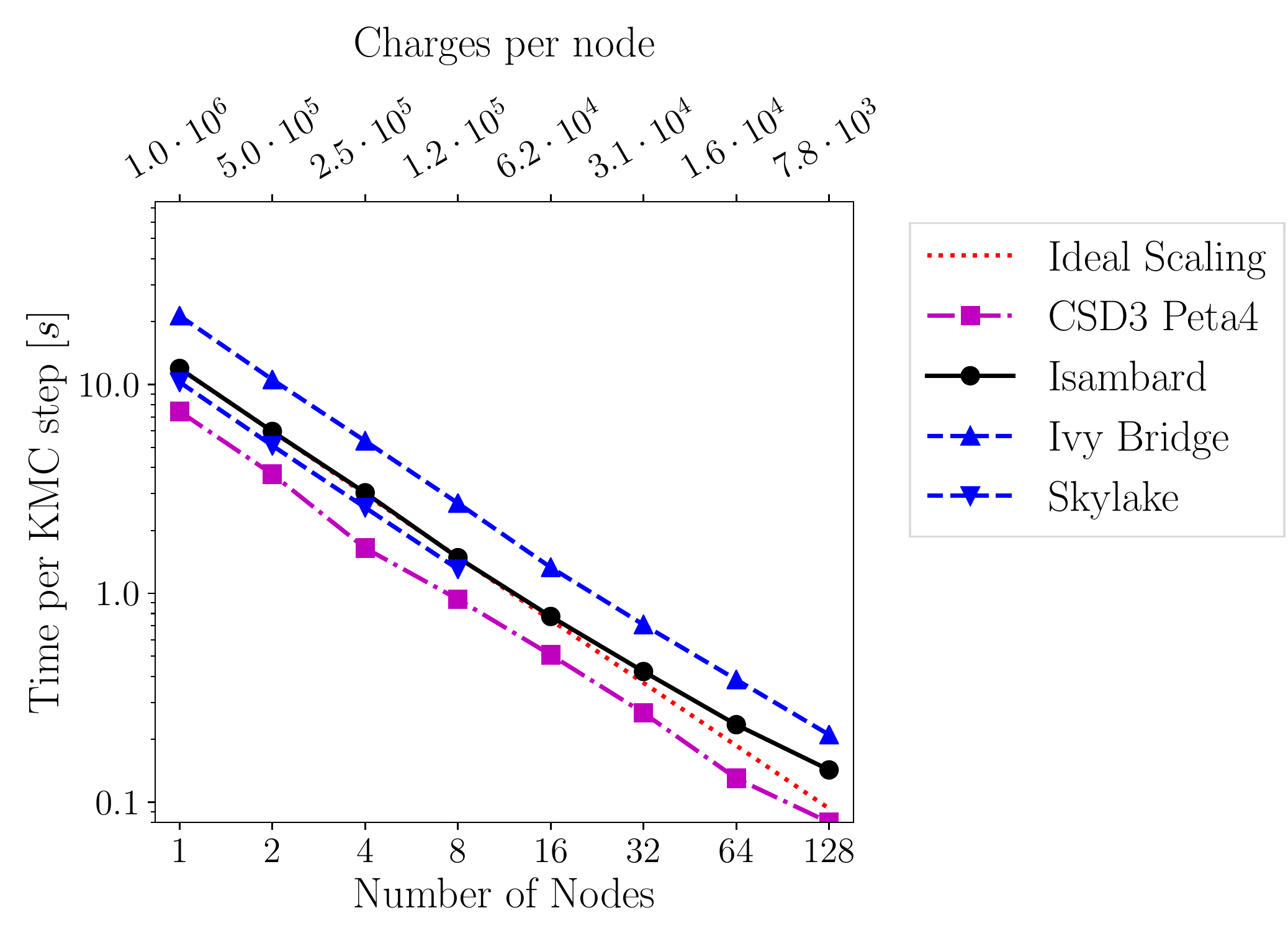}
\end{center}
\end{minipage}
\hfill
\begin{minipage}{0.416\linewidth}
\begin{center}
    \includegraphics[width=\linewidth]{\figdir/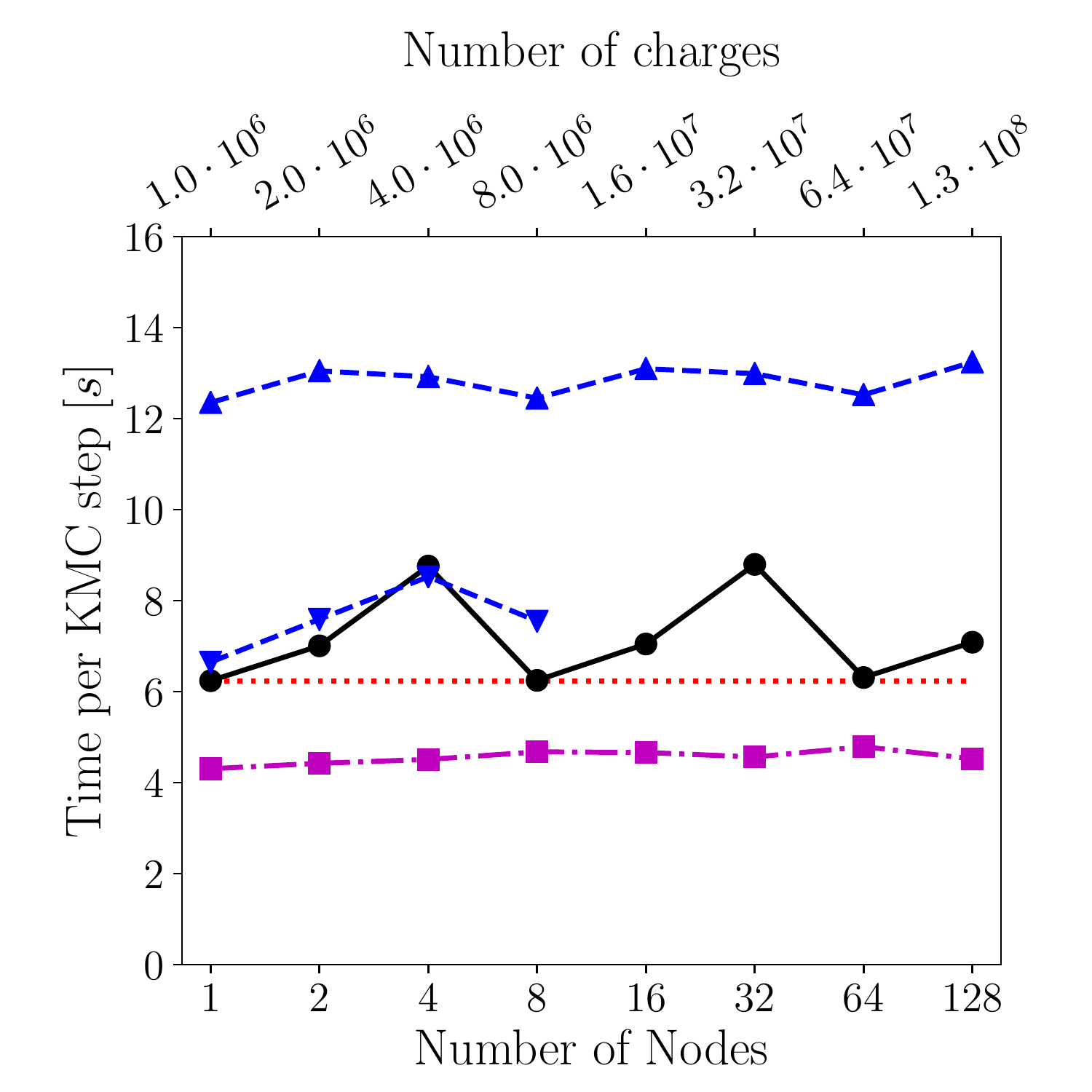}
\end{center}
\end{minipage}
    \caption{Strong (left) and weak (right) scaling experiments. The time per KMC step is plotted against number of compute nodes. For the strong scaling experiment the total number of charges is $N = 10^6$.}
    \label{fig:kmc_strong_weak_scaling}
\end{center}
\end{figure}

The time $t(P,N)$ per KMC step for this model system with $N=10^6$ particles running on $P$ nodes is given in Tab.\ \ref{tab:kmc_strong_scaling} for different machines and also plotted in in Fig.\ \ref{fig:kmc_strong_weak_scaling} (left).
As usual, the parallel efficiency $E_S(P;N)$ for the strong scaling experiment is defined relative to one node:
\begin{equation}
\begin{aligned}
  E_S(P;N) = \frac{t(1,N)}{P\cdot t(P,N)}.
\end{aligned}
\label{eqn:strong_eff}
\end{equation}

For the corresponding weak scaling experiment we fix the \textit{local} problem size, i.e. the number of charges per node, to $N_{\text{local}}=10^6$ and increase the \textit{total} number of charges $N=P\cdot N_{\text{local}}$ in proportion to the number of nodes. Results for $t(P,N)=t(P,P\cdot N_{\text{local}})$ are given in Tab. \ref{tab:kmc_weak_scaling} and Fig. \ref{fig:kmc_strong_weak_scaling} (right), where the parallel efficiency in this weak scaling experiment is defined as
\begin{equation}
  E(P;N_{\text{local}}) = \frac{t(1,N_{\text{local}})}{t(P,P\cdot N_{\text{local}})}.
  \label{eqn:weak_eff}
\end{equation}

\begin{table*}
\begin{center}
  \begin{tabular}{r|rr|rr|rr|rr}
    \hline
      & \multicolumn{8}{c}{Time per KMC step [s]}\\
\cline{2-9}
 $P$          & \multicolumn{2}{c}{CSD3 Peta4} & \multicolumn{2}{c}{Isambard} & \multicolumn{2}{c}{Ivy Bridge} & \multicolumn{2}{c}{Skylake}\\
    \hline
    \hline
   1 &    7.43 & (100.0\%) &    11.93 & (100.0\%) &    21.33 & (100.0\%) &    10.26 & (100.0\%)\\
   2 &    3.72 & (99.9\%) &     5.96 & (100.1\%) &    10.55 & (101.1\%) &     5.09 & (100.8\%)\\
   4 &    1.65 & (112.8\%) &     3.03 & (98.4\%) &     5.36 & (99.5\%) &     2.57 & (99.7\%)\\
   8 &    0.94 & (99.1\%) &     1.48 & (100.7\%) &     2.70 & (98.7\%) &     1.30 & (98.3\%)\\
  16 &    0.51 & (91.5\%) &     0.77 & (96.3\%) &     1.33 & (100.2\%) &  & \\
  32 &    0.27 & (86.8\%) &     0.42 & (88.3\%) &     0.71 & (94.2\%) &  & \\
  64 &    0.13 & (89.3\%) &     0.23 & (79.4\%) &     0.39 & (86.3\%) &  & \\
 128 &    0.08 & (72.5\%) &     0.14 & (65.3\%) &     0.21 & (79.2\%) &  & \\
    \hline
  \end{tabular}
  \caption{Time per KMC step from a strong scaling experiment for $N = 10^6$ charges and increasing numbers of nodes $P$. Parallel $E_S(P;N)$ efficiency as defined by Eq. \eqref{eqn:strong_eff} is given as relative to a single node in brackets.}
\label{tab:kmc_strong_scaling}
\end{center}
\end{table*}

\begin{table*}
\begin{center}
  \begin{tabular}{r|rr|rr|rr|rr}
    \hline
                 & \multicolumn{8}{c}{Time per KMC step [s]}\\
\cline{2-9}
 $P$          & \multicolumn{2}{c}{CSD3 Peta4} & \multicolumn{2}{c}{Isambard} & \multicolumn{2}{c}{Ivy Bridge} & \multicolumn{2}{c}{Skylake}\\
    \hline
    \hline
   1 &    4.30 & (100.0\%) &     6.24 & (100.0\%) &    12.35 & (100.0\%) &     6.65 & (100.0\%)\\
   2 &    4.42 & (97.3\%) &     7.01 & (89.1\%) &    13.05 & (94.7\%) &     7.59 & (87.6\%)\\
   4 &    4.51 & (95.5\%) &     8.76 & (71.2\%) &    12.92 & (95.6\%) &     8.52 & (78.0\%)\\
   8 &    4.68 & (92.1\%) &     6.25 & (99.9\%) &    12.45 & (99.2\%) &     7.55 & (88.1\%)\\
  16 &    4.66 & (92.3\%) &     7.05 & (88.5\%) &    13.10 & (94.3\%) &  & \\
  32 &    4.56 & (94.3\%) &     8.80 & (70.9\%) &    12.99 & (95.1\%) &  & \\
  64 &    4.79 & (89.9\%) &     6.31 & (98.9\%) &    12.52 & (98.6\%) &  & \\
 128 &    4.52 & (95.2\%) &     7.09 & (88.1\%) &    13.24 & (93.3\%) &  & \\
    \hline
  \end{tabular}
  \caption{Weak scaling experiment: time per KMC step against number of nodes $P$ with $N_{\text{local}} = 10^6$ charges per node. Parallel efficiency $E(P;N_{\text{local}})$ as defined by Eq. \eqref{eqn:weak_eff} is given as relative to a single node in brackets.}
\label{tab:kmc_weak_scaling}
\end{center}
\end{table*}
\subsection{KMC Simulation of $\alpha$-NPD}\label{sec:results_alphaNPD}
We next investigate the performance of our KMC algorithm when applied to a physically realistic configuration. The system studied is $\alpha$-NPD doped with F6TCNNQ, a hole transporting organic semiconductor material \cite{aNPD}. The scientific results of the simulations will be discussed in a forthcoming publication \cite{Thompson2019} and here we concentrate on evaluating the runtimes for a given setup. The dopant molecules ionise and release a mobile charge carrier; this creates a fixed negative charge and a mobile positive charge.
The kinetic Monte Carlo algorithm describes the hopping of the holes between different $\alpha$-NPD molecules, the hopping rates are described by Marcus theory and are functions of structural energy, temperature and molecular polarisation as well as electrostatic energy.

The KMC code used the \texttt{propose\_with\_dats()} interface and applied a modified version of Algorithm \ref{alg:kmc_ppmd} for bookkeeping operations.
This modified algorithm only updates the proposed positions $\mathcal{R}^{(i)}$ and associated masks $\mathcal{M}^{(i)}$ if charge $i$ is in the vicinity of the previously accepted move.
This removes redundant bookkeeping operations at each KMC step.
The $\alpha$-NPD simulations were performed with a range of applied voltages to investigate the charge mobility dependence on applied voltage.
For this a constant external electric field in one direction is added to simulate a given applied potential difference across the domain.
The electrostatic field induced by the charges is assumed to be periodic in all three dimensions and the charges were allowed to wrap around the simulation domain upon reaching the boundary\footnote{Note that while this was not done here, our code also allows enforcing Dirichlet boundary conditions at the top and bottom of the domain by using mirror charges, as described in Section \ref{sec:boundary_conditions}}.

\begin{figure}
\begin{center}
    \includegraphics[width=0.450\linewidth]{\figdir/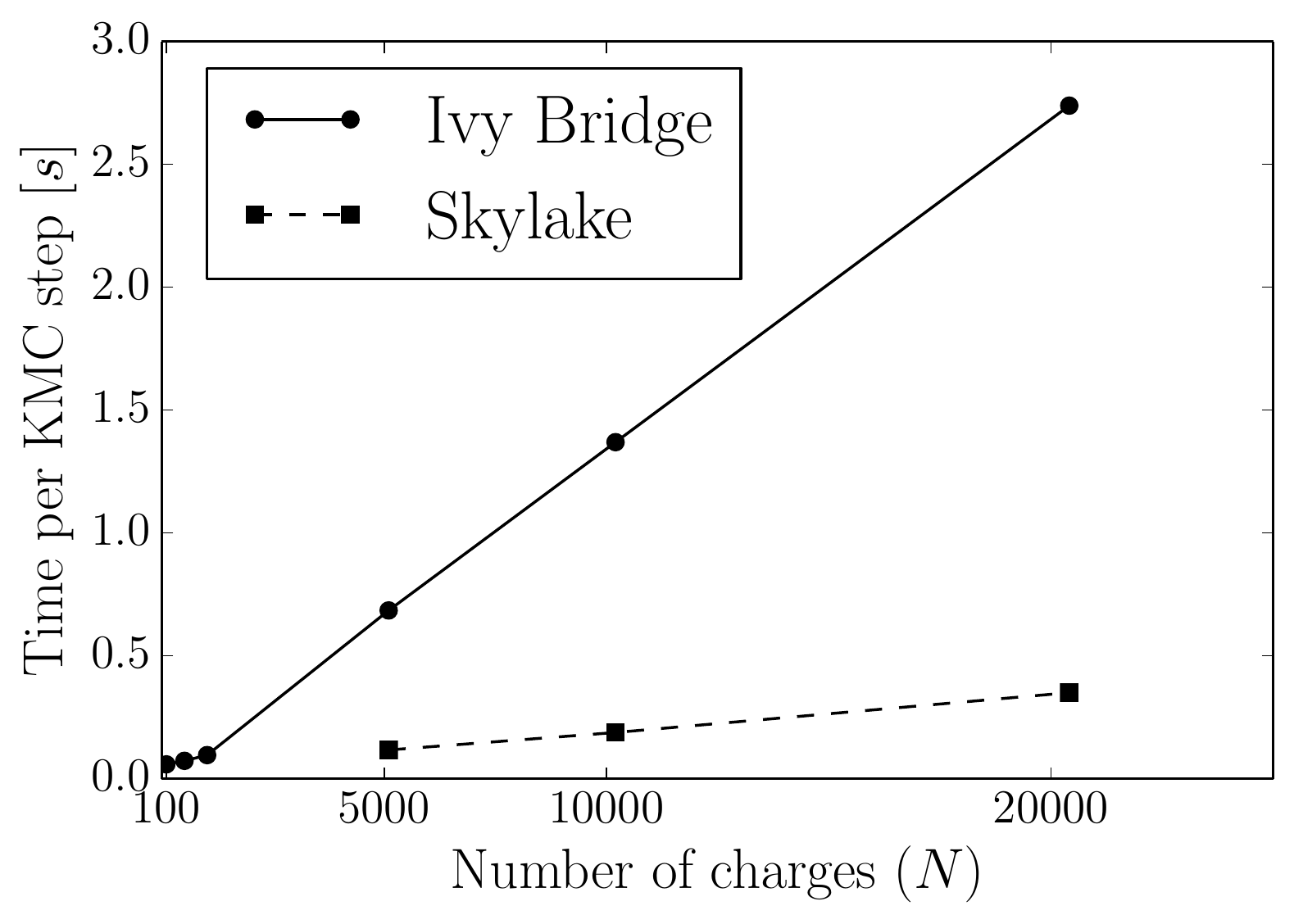}
    \caption{Time per KMC step for $\alpha$-NPD simulations executed on a single node containing either two Ivy Bridge E5-2650v2 or two Skylake Gold 6126 CPUs. The Ivy Bridge simulations used 4 CPU cores and $L=3$ FMM mesh levels. The Skylake simulations used 12 CPU cores and $L=4$ mesh levels.}
    \label{fig:kmc_anpd_timing}
\end{center}
\end{figure}

To exploit additional parallelism between ensembles, multiple instances of the $\alpha$-NPD simulation were run simultaneously. Each instance was executed on either 4 Ivy Bridge cores (running 4 simulations in parallel on a full node) or 12 Skylake cores (2 simulations per node). The dependence of the time per KMC step on the number of charges in the system is shown in Tab. \ref{tab:kmc_anpd_timing} and plotted in Fig. \ref{fig:kmc_anpd_timing}. For the largest studied system with 20412 charges, one KMC step takes 0.35s when run on a full Skylake socket with 12 cores.
\begin{table}[]
    \begin{center}
        \begin{tabular}{rr|rr}
            \hline
                doping & charges & \multicolumn{2}{c}{Time per KMC step [s]} \\
                      \% & ($N$) & Ivy Bridge & Skylake \\
            \hline
            \hline
                0.01 & 102   & 0.058  &      \\
                0.05 & 510   & 0.073  &      \\
                0.1  & 1020  & 0.096  &      \\
                0.5  & 5103  & 0.68   & 0.12 \\
                1.0  & 10206 & 1.37   & 0.19 \\
                2.0  & 20412 & 2.74   & 0.35 \\
            \hline
        \end{tabular}
        \caption{Time per KMC step for $\alpha$-NPD simulations executed on a single node containing either two Ivy Bridge E5-2650v2 or two Skylake Gold 6126 CPUs. The Ivy Bridge simulations used 4 CPU cores and $L=3$ FMM mesh levels. The Skylake simulations used 12 CPU cores and $L=4$ mesh levels.}
        \label{tab:kmc_anpd_timing}
    \end{center}
\end{table}

\subsection{Performance comparison to ScalFMM}
Finally, we verify that our implementation in the PPMD framework is indeed efficient. There is currently no existing library which implements the bespoke FMM method for KMC simulations developed in this paper. A fair comparison to other KMC packages based on Ewald summation or approximations described in \cite{vanderHolst2011,Kordt2015} would have to be carried out at a fixed, problem dependent error tolerance and is therefore more appropriate for future application-driven studies. A meaningful intercomparison study might also be outright impossible since most of the methods in the literature introduce uncontrollable errors. To nevertheless assess the absolute performance of our implementation we compare the runtime of the \textit{standard} FMM method in Algorithms \ref{alg:fmm} and \ref{alg:fmm_evaluation} to the freely available ScalFMM library \cite{Agullo2014}. Given that no extensive attempts have been made to optimise our code, the aim of this comparison is to verify that the performance is in the same ballpark as a published reference implementation.

For this, we measured the time spent in one iteration of a Velocity Verlet integrator for a set of particles which interact via a Coulomb potential. Although the setup of the system differed slightly (the ScalFMM test case uses free-space boundary conditions and a pure $\sim 1/r$ repulsive potential, whereas our code was used to simulate a charge-neutral system with periodic boundary conditions and an additional short-range Lennard-Jones interaction), we believe this allows a sensible assessment of the relative performance of the codes. The results in Tab. \ref{tab:performance_comparison_ScalFMM} confirm that in both cases the runtime is dominated by the FMM algorithm. To allow a fair comparison, the parameters were tuned such that both simulations are carried out at comparable accuracy. The ScalFMM code uses the uniform Lagrange interpolation kernel, which depends on the number $\ell_{\text{Lag}}$ of terms. As argued in \cite{Agullo2014} and confirmed in by the ScalFMM developers \cite{BramasCoulaud2019}, setting $\ell_{\text{Lag}}=5$ is expected to give a relative accuracy of around $\epsilon=10^{-5}$ in energy and force calculations. As shown in \cite{Saunders2018a}, the same relative accuracy is achieved by using $p=10$ multipole expansion terms in our code. 
All runs were carried out on a single, fully utilised, Ivy Bridge node. While the ScalFMM code was run in OpenMP mode with 16 threads, results on both a pure MPI run and a mixed mode MPI/OpenMP configuration (2 MPI processes with 8 threads each) are reported for our code.
Tab. \ref{tab:performance_comparison_ScalFMM} shows measured runtimes for $N=10^6$ and $N=4\cdot10^6$ particles. For our implementation the optimal number of levels turned out to be $L=5$ for $N=10^6$ and $L=6$ for $N=4\cdot10^6$, whereas the ScalFMM code gave the best results if the equivalent octree depth was set to $6$ in both cases.
\begin{table*}
\begin{center}
  \begin{tabular}{cc|rrr|rrr}
    & \# particles & \multicolumn{3}{c|}{$N=10^6$} & \multicolumn{3}{c}{$N=4\cdot10^6$} \\
    implementation & & VV-step & FMM &
    & VV-step & FMM & \\
    \hline\hline
    \multirow{2}{*}{PPMD}   & (MPI only)    & 3.03 & 2.91 & [96\%]
                            & 9.78 & 8.96 & [92\%] \\
    &  (MPI+OpenMP)         & 3.96 & 3.78 & [95\%]
                            & 12.39 & 11.48 & [93\%] \\\hline
    ScalFMM & (OpenMP only) & 1.16 & 1.14 & [99\%]
                            & 4.00 & 3.93 & [98\%] \\
  \end{tabular}
  \caption{Performance comparison of our implementation of the Standard FMM algorithm to ScalFMM. The total time per Velocity Verlet (VV) step is listed in the first column and includes overheads from updating the velocities and positions with axpy operations. For the PPMD implementation it also includes the cost of the short range Lennard-Jones interaction. For each case, the cumulative time spent in the FMM setup phase (Algorithm \ref{alg:fmm}) and the evaluation phase (Algorithm \ref{alg:fmm_evaluation}) is given, together with the percentage of time spent in the FMM calculation. All times are measured in seconds.}
\label{tab:performance_comparison_ScalFMM}
\end{center}
\end{table*}
The results in Tab. \ref{tab:performance_comparison_ScalFMM} confirm that the performance of our code is of the same order of magnitude as the highly optimised ScalFMM library. In fact, the pure MPI implementation is only around $3\times$ slower, which is acceptable given that we have not yet considered further optimisation. Further results on the parallel scalability of our standard FMM implementation can be found in \secapp\ref{sec:results_parallelFMM}.
\section{Conclusions}\label{sec:conclusions}
In this paper we described how the Fast Multipole Method can be adapted to allow the efficient and accurate treatment of electrostatic interactions in kinetic Monte Carlo simulation. Although a recent publication \cite{Li2018} claimed that this would not be possible, we demonstrated that our algorithm scales linearly with the number of charges. This was confirmed numerically by measuring the time $t_{\text{step}}$ per KMC step for systems with up to $1.3 \cdot 10^8$ charges. Running in parallel on $8192$ cores we find $t_{\text{step}} = 7.09$s. We also presented results for a physically relevant $\alpha$-NPD test case with $20412$ particles which could be simulated at a rate of $0.35\text{s}$ per KMC step on a single 12-core Skylake CPU. 

By facilitating the simulation of much larger systems with realistic electrostatics, our new library will allow step-changes in the KMC simulation of energy materials. While the focus of the paper is on the description of the algorithm, its implementation and parallel performance results, a forthcoming publication \cite{Thompson2019} discusses further results for physically relevant systems.

Our code provides an intuitive user interface and we showed that code generation techniques guarantee excellent scalability and performance on modern HPC installations. In principle the code is performance portable, and has so far been implemented for CPU chips, using hybrid MPI+OpenMP parallelisation. A GPU backend is currently being developed.

An interesting line of future work will be to combine our improved FMM algorithm with novel KMC approaches, such as multilevel KMC techniques. Those methods allow more efficient simulations by skipping physically less interesting transitions, such as ``rattling'' the frequent repeated hops between pairs of states.

By making suitable adaptations to the FMM algorithm, it will be also possible to reduce the cost of \textit{standard} Monte Carlo (MC) simulations. Similar improvements have already been studied for Ewald summation \cite{Ewald1921,Frenkel2001}, for which the change in electrostatic energy \textit{per MC move} can be calculated at a computational complexity of $\mathcal{O}(\sqrt{N})$. This is because the overall $\mathcal{O}(N^{3/2})$ cost of the Ewald-based energy calculation is made up by an iteration over all $N$ particles and a sum over $O(\sqrt{N})$ reciprocal vectors (long-range contribution) and neighbouring particles (short-range contribution). If only $\mathcal{O}(1)$ particles move at each MC step, only a small number of the $\mathcal{O}(\sqrt{N})$ sums have to be evaluated. A similar approach is currently explored in the DL\_MONTE code \cite{Purton2013}, though the implementation at present is $O(N)$. We believe that a suitably modified FMM algorithm will improve on this and limit the computational complexity per individual MC move to $\mathcal{O}(\log(N))$. The key idea is to store the local expansion $\Psi_{\ell,i}$ on each level of the grid hierarchy, instead of accumulating it on the finest level in the downward sweep. While calculation of the electrostatic field requires evaluation of the $\Psi_{\ell,i}$ in one cell on each of the $L\sim \log(N)$ levels, updating the field only requires changes to a constant number of cells per level. Overall, both operations can be carried out in $\mathcal{O}(L)=\mathcal{O}(\log(N))$ time per MC update.
\section*{Acknowledgements}
This research made use of the Balena High Performance Computing (HPC) Service at the University of Bath and the Isambard UK National Tier-2 HPC Service (\url{http://gw4.ac.uk/isambard/}). Isambard is operated by GW4 and the UK Met Office, and funded by EPSRC (EP/P020224/1).
This work was performed using resources provided by the Cambridge Service for Data Driven Discovery (CSD3) operated by the University of Cambridge Research Computing Service (www.csd3.cam.ac.uk), provided by Dell EMC and Intel using Tier-2 funding from the Engineering and Physical Sciences Research Council (capital grant EP/P020259/1), and DiRAC funding from the Science and Technology Facilities Council (www.dirac.ac.uk).
This project has received funding from the European Union’s Horizon 2020 research and innovation programme under grant agreements No 646176 and No 824158. William Saunders was funded by an EPSRC studentship during his PhD. We would like to thank B\'{e}renger Bramas and Olivier Coulaud for help with running ScalFMM on our cluster.
\appendix
\section{Standard FMM algorithm}\label{sec:FMM_algorithm_details}
For reference the standard FMM algorithm described in Section \ref{sec:FMM} is written down in Algorithms \ref{alg:fmm} and \ref{alg:fmm_evaluation}. In addition to the expansions $\Psi_{\ell,\cellidxi}$ and $\Phi_{\ell,\cellidxi}$, this requires another $p$-term local expansion $\bar{\Psi}_{\ell,\cellidxi}$ about the centre of cell $\cellidxi$ on level $\ell$. $\bar{\Psi}_{\ell,\cellidxi}$ describes the potential induced by all charges outside the parent of the cell and outside the 26 nearest neighbours of this parent. Algorithm \ref{alg:fmm} also uses the notion of an \textit{interaction list} ($\operatorname{IL}$) of a particular cell. This is important to recursively include contributions from finer levels (see line 18 in Algorithm \ref{alg:fmm}). For a cell $\cellidxi$ on level $\ell$ the interaction list $\text{IL}(\cellidxi)$ is the set of cells which are the children of the parent cell of $\cellidxi$ and its nearest neighbours, but which are well separated from $\cellidxi$, i.e. not direct neighbours of $\cellidxi$ on level $\ell$. Explicitly, the interaction list is given as
\begin{equation*}
\operatorname{IL}(\cellidxi) = \text{children}\left(\mathcal{N}_b\left(\text{parent}(\cellidxi)\right)\right)\backslash\left(\cellidxi\cup\mathcal{N}_b(\cellidxi)\right).
\end{equation*}
On a particular level $\ell$ the local representation of the operators $\mathcal{T}_{\operatorname{MM}}$ for converting between multipole expansions between cells $\cellidxi$ and $\cellidxj$ is given by $\mathcal{T}_{\operatorname{MM}}^{(\ell;\cellidxi,\cellidxj)}$ with corresponding notation for $\mathcal{T}_{\operatorname{ML}}$ and $\mathcal{T}_{\operatorname{LL}}$.

As can be seen in Algorithm \ref{alg:fmm_evaluation}, FMM provides constant-time access to the local expansions (and nearest-neighbour interactions) on the finest level. This property is crucial for the construction of the algorithmically optimal FMM method for KMC simulations presented in this paper.  
\begin{algorithm}
\caption{Fast Multipole Algorithm I. Construct local expansion $\Psi_{L,\cellidxi}$ of long range contribution.}
\label{alg:fmm}
\begin{algorithmic}[1]
  \FORALL{cells $\cellidxi=1,\dots,M=8^{L-1}$}
  \STATE{Construct multipole expansion $\Phi_{L,\cellidxi}$ of all}
  \STATEx{$\quad$charges contained in cell $\cellidxi$}
  \ENDFOR
  \STATEx{\textit{Upward pass:}}
  \FORALL{levels $\ell=L-1,\dots,1$}
  \FORALL{cells $\cellidxi=1,\dots,M_\ell=8^{\ell-1}$}
    \STATE{Set $\Phi_{\ell,\cellidxi}=0$}
    \FORALL{cells $\cellidxj\in\text{children}(\cellidxi)$}
    \STATE{$\Phi_{\ell,\cellidxi}\,\mapsfrom\Phi_{\ell,\cellidxi}+\mathcal{T}^{(\ell;\cellidxi,\cellidxj)}_{\text{MM}}\Phi_{\ell+1,\cellidxj}$}
    \ENDFOR
    \ENDFOR
    \ENDFOR
    \STATEx{\textit{Downward pass:}}
    \FORALL{level $\ell=2,\dots,L$}
    \FORALL{cells $\cellidxi=1,\dots,M_\ell=8^{\ell-1}$}
    \STATE{$\overline{\Psi}_{\ell,\cellidxi}\mapsfrom\mathcal{T}^{(\ell;\cellidxi,\cellidxj)}_{\text{LL}}\Psi_{\ell-1,\cellidxj}$ for $\cellidxj=\operatorname{parent}(\cellidxi)$}
    \ENDFOR
    \FORALL{cells $\cellidxi=1,\dots,M_\ell=8^{\ell-1}$}
    \STATE{$\Psi_{\ell,\cellidxi}\mapsfrom\overline{\Psi}_{\ell,\cellidxi}$}
    \FORALL{cells $\cellidxj\in\text{IL}(\cellidxi)$}
    \STATE{$\Psi_{\ell,\cellidxi}\mapsfrom\Psi_{\ell,\cellidxi}+\mathcal{T}^{(\ell;\cellidxi,\cellidxj)}_{\text{ML}}\Phi_{\ell,\cellidxj}$}
    \ENDFOR
    \ENDFOR
    \ENDFOR
\end{algorithmic}
\end{algorithm}
\begin{algorithm}
\caption{Fast Multipole Algorithm II. Evaluate long- and short- range contributions to potential energy $U$.}
\label{alg:fmm_evaluation}
\begin{algorithmic}[1]
  \STATE{Set $U=0$}
  \FORALL{cells $\cellidxi=1,\dots,M$}
  \FORALL{charges $\partidxi$ in cell $\cellidxi$}
  \STATE{$U \mapsfrom U + \frac{1}{2}\Psi_{L,\cellidxi}\left(\vec{r}^{(\partidxi)}\right)$}
  \FORALL{charges $\partidxj\ne\partidxi$ in $\cellidxi\cup \mathcal{N}_b(\cellidxi)$}
  \STATE{$U \mapsfrom U + \frac{1}{2}q^{(\partidxi)}q^{(\partidxj)}/\left|\vec{r}^{(\partidxi)}-\vec{r}^{(\partidxj)}\right|$}
  \ENDFOR
  \ENDFOR
  \ENDFOR
\end{algorithmic}
\end{algorithm}
\section{Dipole correction}\label{sec:dipole_correction_details}
\begin{figure}
\begin{center}
\includegraphics[width=0.45\linewidth]{\figdir/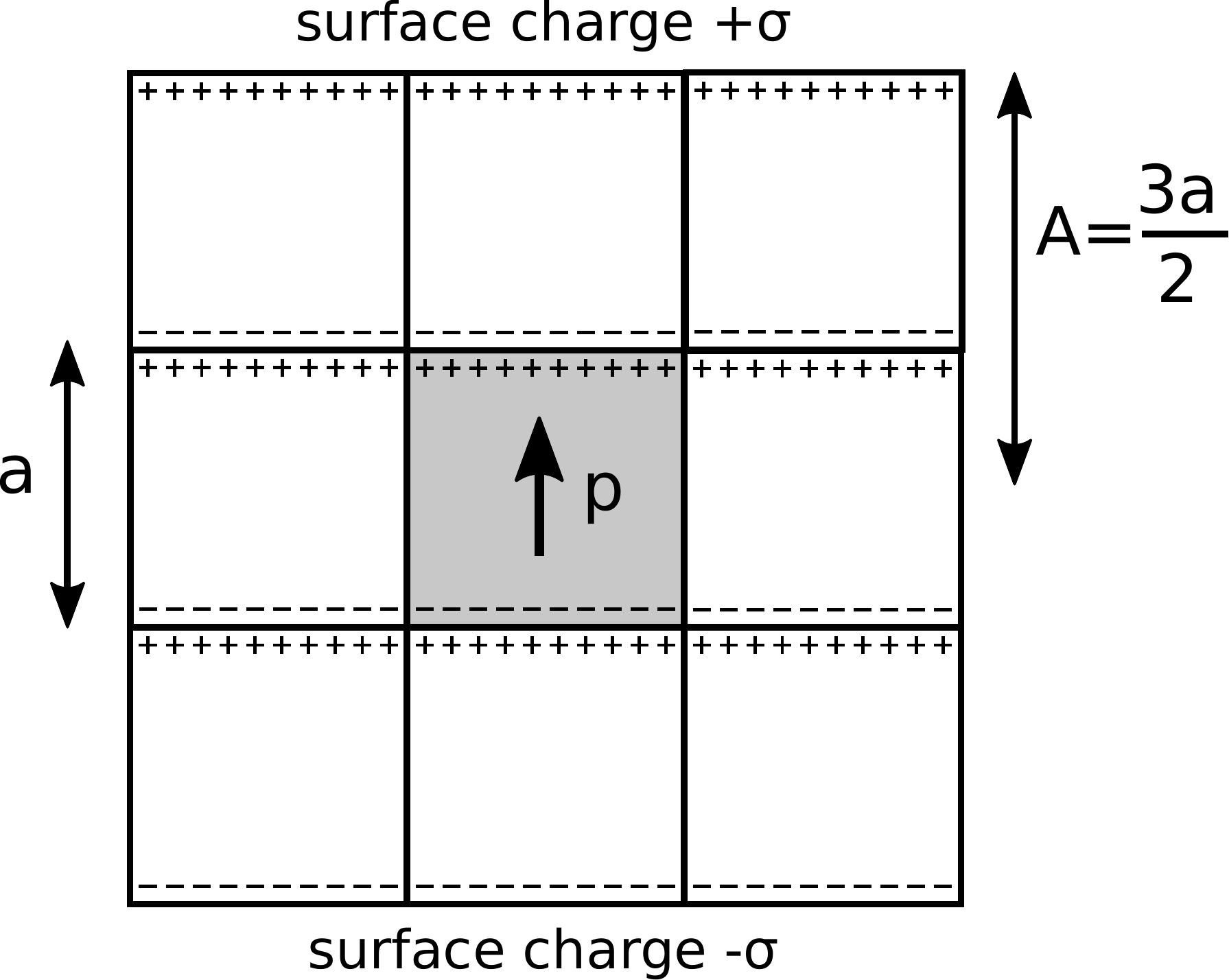}
\caption{Charge distribution for working out dipole correction. The primary image is shown in gray. Note that the charges cancel at all interior interfaces, leaving surface charges at the top- and bottom boundaries.}
\label{fig:dipole_correction_details}
\end{center}
\end{figure}

As discussed in Section \ref{sec:dipole_correction}, care has to be taken if the dipole moment of the charge distribution is non-zero for periodic- or Dirichlet boundary conditions. Here we derive the correction term which needs to be added to the electric field to enforce the zero surface-charge boundary condition at infinity.
First observe that due to the nature of the multipole expansion and the linearity of electrostatics, it is sufficient to derive the term for one particular charge configuration with a given dipole moment. Assume that there is a surface charge density of $+\sigma$ on the top face of the domain, and an opposite density of $-\sigma$ at the opposite face (see Fig. \ref{fig:dipole_correction_details}).
This induces a dipole moment per unit volume of $p=\sigma$. Let $\phi^{(\sr)}$ be the potential which is induced by the primary cell and its 26 neighbours, i.e. the near-field contribution. The corresponding far-field contribution is denoted by $\phi^{(\lr)}$. Calculation of $\phi^{(\sr)}$ at the point $\vec{r}=(x,y,z)$ is straightforward, since we only need to compute the potential generated by two oppositely charged plates of size $(3/2a)^2$ at a distance of $A=3/2a$ from the origin. The fundamental solution of the Poisson equation and the superposition principle gives
\begin{equation*}
  \begin{aligned}
  \phi^{(\sr)} &= \int_{-A}^{A}\int_{-A}^{A}
  \frac{\sigma\;d\eta\;d\xi}
       {\sqrt{(x-\eta)^2+(y-\xi)^2+(A-z)^2}}
       - \int_{-A}^{A}\int_{-A}^{A}
  \frac{\sigma\;d\eta\;d\xi}
  {\sqrt{(x-\eta)^2+(y-\xi)^2+(A+z)^2}}
  \end{aligned}
\end{equation*}
We want to calculate the electric field $E$ at the origin, which is given by Taylor-expanding the total potential
\begin{equation*}
  \phi(\vec{r}) = \phi^{(\sr)}(\vec{r}) + \phi^{(\lr)}(\vec{r}) = -Ez + \dots.
\end{equation*}
(note that due to symmetry, for this particular setup there are no contributions which are linear in $x$ or $y$). Since the coefficient $R_2^m = 0$ of the matrix $\mathcal{R}$ introduced in \cite{Amisaki2000} is zero, there is no contribution to $E$ from $\phi^{(\lr)}$. This implies that $E$ can be obtained by taking the derivative of $\phi^{(\sr)}$ at the origin. The resulting surface integral is readily evaluated to obtain
\begin{equation*}
  \begin{aligned}
    E &= -\frac{\partial\phi^{(\sr)}}{\partial z}\vert_{\vec{r}=0}
    = -2A\sigma \int_{-A}^{A}\int_{-A}^{A} \left(A^2+\eta^2+\xi^2\right)^{-3/2}\;d\eta\;d\xi
    = -\frac{4\pi}{3}\sigma.
  \end{aligned}
\end{equation*}
The same argument can be applied for dipoles pointing in the $x$-  and $y$- direction; recalling that $\sigma=p$, this implies that the field of a vector-valued dipole density $\vec{p}$ is given by
\begin{equation*}
  \vec{E} = -\frac{4\pi}{3}\vec{p}.
\end{equation*}
\section{Improved interface for proposals}\label{sec:improved_proposal_interface}
While the \texttt{propose()} method described in Section \ref{sec:proposing_moves} is intuitive and easy to use, it is not optimal in terms of efficiency. This is because the tuple of proposed moves $\mathcal{P}$ in Eq. \eqref{eqn:tuple_P} has to be converted to an internal data structure before the moves can be passed to our C implementation. To overcome this issue we provide an alternative interface which is more efficient, but requires additional work from the user since the corresponding \texttt{propose\_with\_dats()} method expects the data to be given in a particular, structured format: the proposed moves have to be encoded as a set of particle properties which are stored in \texttt{ParticleDat} instances. In contrast to the \texttt{propose()} interface, which can operate on a subset of particles, potential new positions have to be specified for all particles in the system. For this, the particles are separated into $M$ different types, such that a particle of type $t\in \{1,\dots,M\}$ can potentially transition to $c_t\in \mathbb{N}_0$ new positions. Note that $c_t$ can be zero, and particles of this type are not able to move at all. The type of a particle could for example depend on the topology of the lattice or the local environment of the lattice site it currently occupies. In this case the type can change during the simulation; an example is described in \secapp\ref{sec:KMC_bookkeeping}. The set $\mathcal{C} = \{c_1,c_2,\dots,c_M\}$ is represented by a $M$-dimensional \texttt{ScalarArray}. The types of all particles are stored in an integer-valued \texttt{ParticleDat} $\mathcal{T}$, such that $\mathcal{T}^{(\partidxi)}\in \{1,\dots,M\}$ is the type of particle $\partidxi$, which can therefore hop to $c_{\mathcal{T}^{(\partidxi)}}$ potential new sites. Let $K = \max\left( \mathcal{C} \right)$ be the maximum number of moves at any site. An additional real-valued \texttt{ParticleDat} $\mathcal{R}$ with $3K$ components is used to store the locations of the potential destinations of all particles. For particle $\partidxi$ the entry $\mathcal{R}^{(\partidxi)}$ contains the list $\{{\vec{r}'}_1^{(\partidxi)},{\vec{r}'}_2^{(\partidxi)},\dots,{\vec{r}'}_K^{(\partidxi)}\}$ of three dimensional vectors where any entries ${\vec{r}'}_k^{(\partidxi)}$ with $k>c_{\mathcal{T}^{(\partidxi)}}$ are irrelevant and may contain arbitrary values. Depending on the local environment of a particle and other particles in its vicinity, particular hops might be blocked for a particular configuration. To avoid changing the type of those particles and shuffling around the entries of $\mathcal{R}^{(\partidxi)}$ to account for this, certain transitions can be marked as ``forbidden'' by setting a flag in a separate \texttt{ParticleDat} $\mathcal{M}$ with $K$ components. Each entry $\mathcal{M}^{(\partidxi)}$ contains a list $\{m_1^{(\partidxi)},m_2^{(\partidxi)},\dotsc,m_K^{(\partidxi)}\}$ where the flags $m_k^{(\partidxi)}=1$ signals that the transition ${\vec{r}'}_k^{(\partidxi)}\leftarrow\vec{r}^{(\partidxi)}$ for particle $\partidxi$ is allowed. If $m_k^{(\partidxi)}<1$ this transition is forbidden and will not be considered in the calculation of energy changes for the proposed moves.

The \texttt{ScalarArray} $\mathcal{C}$ and the \texttt{ParticleDat}s $\mathcal{T}$, $\mathcal{R}$, $\mathcal{M}$ and $\mathcal{U}$ are passed as inputs to the \texttt{propose\_with\_dats()} method, which populates the \texttt{ParticleDat} $\mathcal{U}$ of $K$ components with the energy changes of all proposed moves.
More specifically, the entry for particle $\partidxi$ is the list $\mathcal{U}^{(\partidxi)}=\{\Delta U_1^{(\partidxi)},\Delta U_2^{(\partidxi)},\dotsc,\Delta U_K^{(\partidxi)}\}$ such that $\Delta U_k^{(\partidxi)}$ contains the change in electrostatic energy for the move ${\vec{r}'}_k^{(\partidxi)}\leftarrow \vec{r}^{(\partidxi)}$.
Entries for moves which are marked as forbidden through the mask $\mathcal{M}^{(\partidxi)}$ or for which $k>c_{\mathcal{T}^{(\partidxi)}}$ are undefined.

\begin{figure}
\begin{center}
  \includegraphics[width=0.5\linewidth]{\figdir/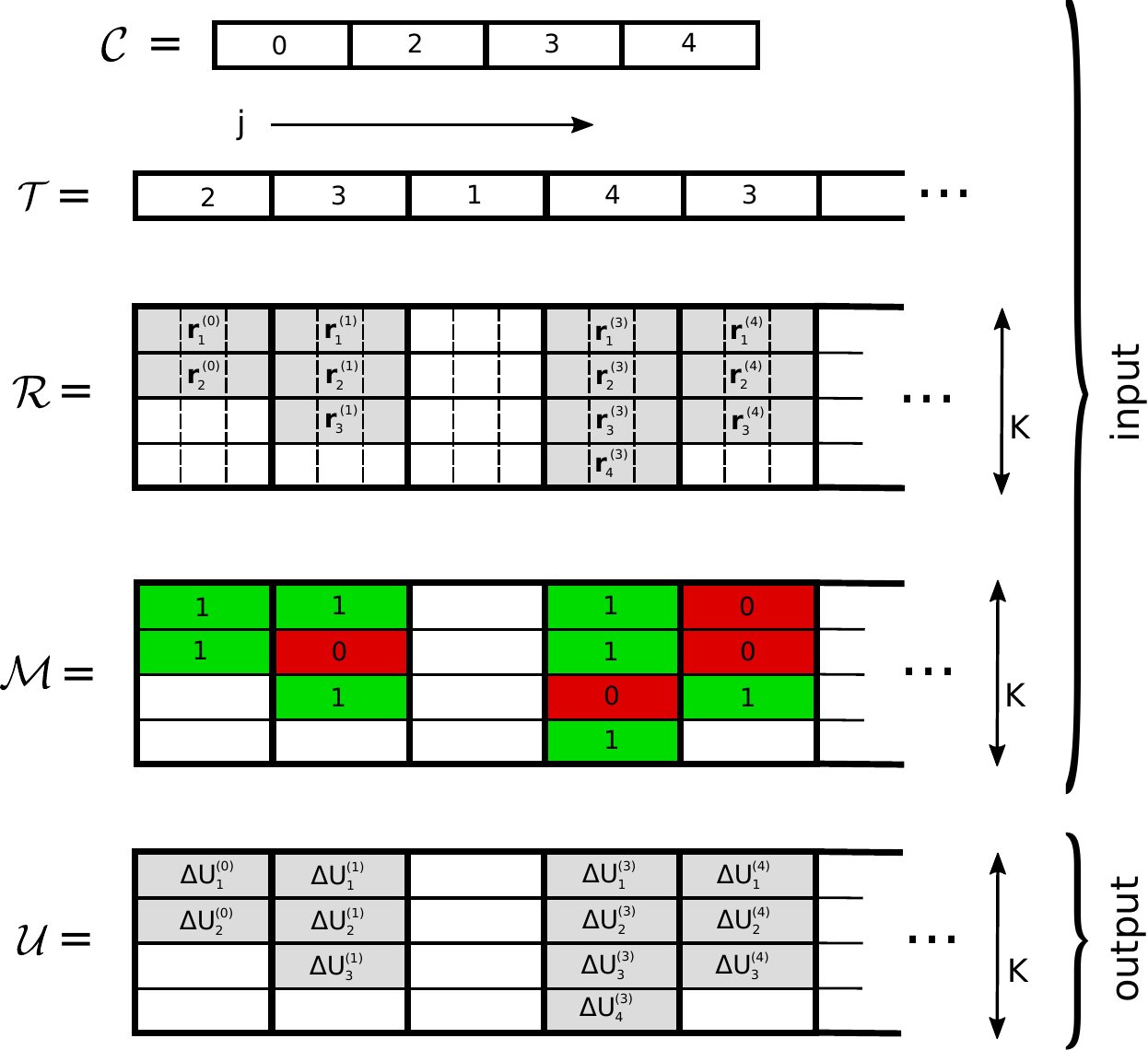}
\caption{Data structures (\texttt{ScalarArray} $\mathcal{C}$ and \texttt{ParticleDat}s $\mathcal{T}$, $\mathcal{R}$, $\mathcal{M}$ and $\mathcal{U}$) used by the \texttt{propose\_with\_dats()} interface for proposing potential moves.}
\label{fig:propose_with_dats}
\end{center}
\end{figure}

An example of the \texttt{ParticleDat}s $\mathcal{T}$, $\mathcal{R}$, $\mathcal{M}$ and $\mathcal{U}$ is shown in Fig. \ref{fig:propose_with_dats}. In this case each particle can hop to between two and four sites, or not hop at all. The set which described the four different types of particles is therefore given by $\mathcal{C}=\{0,2,3,4\}$ with $K=\max(\mathcal{C})=4$.

Using the \texttt{propose\_with\_dats()} and data structures provided by PPMD allows the entire KMC implementation to be written in the looping operations of PPMD. Apart from guaranteeing the efficiency of code, the user never has to explicitly insert parallelisation calls. To illustrate how this can be done, we now show how the inputs to the \texttt{propose\_with\_dats()} method can be set up in PPMD for a particular use case.
\subsection{Selection of allowed moves in PPMD}\label{sec:KMC_bookkeeping}
\begin{figure}
\begin{center}
\includegraphics[width=0.35\linewidth]{\figdir/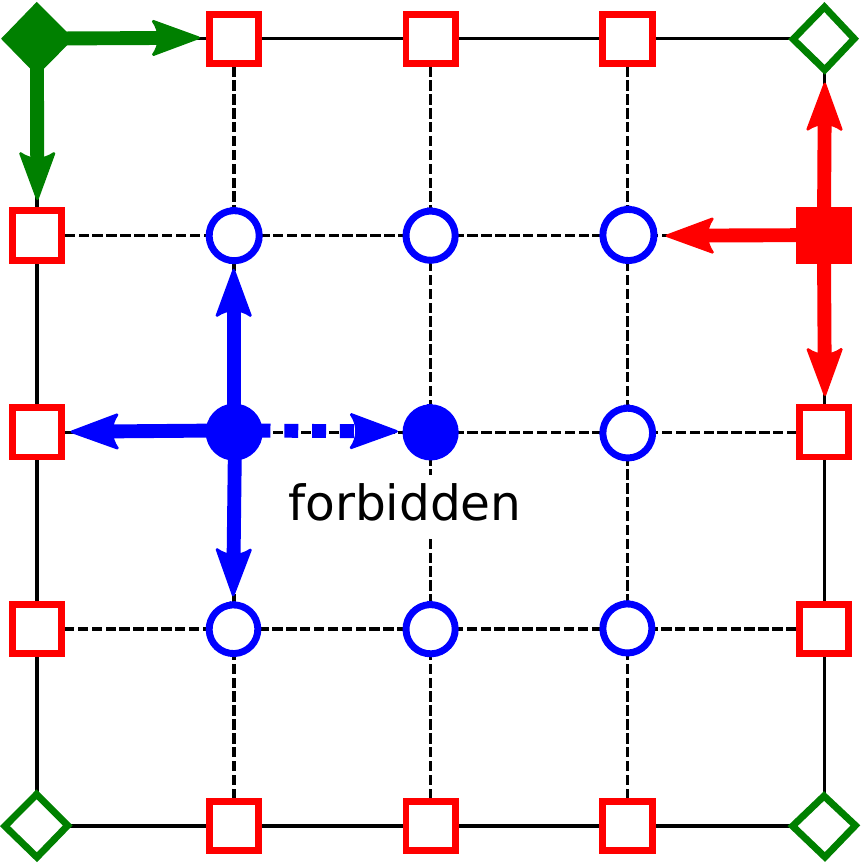}
\caption{Two dimensional grid used for the example in \secapp\ref{sec:KMC_bookkeeping}. Sites of different categories (interior, edge, corner) are marked by different symbols and occupied sites are shown as filled.}
\label{fig:grid_example}
\end{center}
\end{figure}

For this example we assume that the system consists of charges which can hop between the sites of a regular two-dimensional lattice $\Lambda$ with spacing $h$ embedded in three dimensional space
\begin{equation*}
\Lambda = \{\vec{x}=h\vec{n}:\vec{n}\in\mathbb{Z}^2\times \{0\},|\vec{x}|<\frac{1}{2}a\}
\end{equation*}
Recall that the simulation domain is a box of width $a$, and periodic boundary conditions are assumed for the electrostatic potential. However, we assume that the charges can not hop across the domain boundary. The total number of charges $N$ is assumed to be much smaller than the total number of sites and each site can be occupied by at most one charge. In this example we further assume that charges can only hop to directly neighbouring sites. Note that the number of sites a charge can hop to, i.e. its type, depends on whether it is in the interior of the domain or on the boundary, see Fig. \ref{fig:grid_example}.
When setting up the input for \texttt{propose\_with\_dats()}, the following points have to be taken into account:
\begin{itemize}
\item All charges need to be be assigned a type by setting the entries in the \texttt{ParticleDat} $\mathcal{T}$. This depends on the lattice site the particle currently occupies.
\item The potential destinations for all particles have to be worked out in each KMC step by populating the \texttt{ParticleDat} $\mathcal{R}$.
\item Potential hops to already occupied sites need to be masked by setting the entries of the \texttt{ParticleDat} $\mathcal{M}$; again this has to be done in each KMC step.
\end{itemize}
The sites can be arranged into nine different categories (one interior, one for every outer edge/corner of the domain, see Fig. \ref{fig:grid_example}). Since particles sites in the interior of the domain have four direct neighbours, sites on the edges have four and sites in the corners only two, we set
\begin{equation*}
  \mathcal{C}=\{c_1,c_2,\dotsc,c_9\} = \{4,3,3,3,3,2,2,2,2\}.
\end{equation*}
Further, for each site category $t\in\{1,\dots,9\}$ we define an ordered set of integer offsets
\begin{equation*}
  \mathcal{D}^{(t)} = \{\vec{\delta}^{(t)}_1,\vec{\delta}^{(t)}_2,\dotsc,\vec{\delta}^{(t)}_{c_t}\}\subset \mathbb{Z}^3
\end{equation*}
which describes the potential relative hops of a particle located at a site of this category in units of the lattice spacing,
\begin{equation*}
\begin{aligned}
  \mathcal{D}^{(1)} &= \{ (+1,0,0),(-1,0,0),(0,+1,0),(0,-1,0)\},\\
  \mathcal{D}^{(2)} &= \{ (+1,0,0),(-1,0,0),(0,+1,0)\},\\
  \mathcal{D}^{(3)} &= \{ (+1,0,0),(-1,0,0),(0,-1,0)\},\\
\dots
\end{aligned}
\end{equation*}
Now consider the charge with index $\partidxi$, which is of type $t=\mathcal{T}^{(\partidxi)}$ and currently located at position $\vec{r}^{(\partidxi)}\in\Lambda$. Since the number of potential destinations is $c_t$, this charge can potentially hop to any point in the set
\begin{equation*}
  \mathcal{R}^{(\partidxi)} = \{\vec{r}^{(\partidxi)}+h\vec{\delta}^{(t)}_1,\vec{r}^{(\partidxi)}+h\vec{\delta}^{(t)}_2,\dotsc,\vec{r}^{(\partidxi)}+h\vec{\delta}^{(t)}_{c_t}\} \subset \Lambda.
\end{equation*}
This can be implemented by updating the entries $\mathcal{R}^{(\partidxi)}$ of the \texttt{ParticleDat} $\mathcal{R}$ with a \texttt{ParticleLoop} in the PPMD code.

Finally, forbidden moves to already occupied sites have to be masked by setting appropriate flags in $\mathcal{M}$. This is done by considering all pairs $(\partidxi,\partidxj)$ of particles and setting the entry $m_k^{(\partidxi)}$ of $\mathcal{M}^{(\partidxi)}$ to zero if the $k$-th entry of $\mathcal{R}^{(\partidxi)}$ is identical to the current position $\vec{r}^{(\partidxj)}$ of the other particle in the pair, i.e. if $|{\vec{r}'}^{(\partidxi)}_k-\vec{r}^{(\partidxj)}|=0$. In PPMD this operation can be realised with a \texttt{PairLoop}.

The pseudocode in Algorithm \ref{alg:kmc_ppmd} provides an overview of the PPMD implementation of the book-keeping operations discussed in this section. To set the types of the particles it is assumed that there is a function $T$ which returns the category of a lattice site located at $\vec{r}$.
\begin{algorithm}
    \caption{Overview of bookkeeping operations for updating the \texttt{ParticleDat}s $\mathcal{T}$, $\mathcal{R}$ and $\mathcal{M}$ in each KMC step, as described in \secapp\ref{sec:KMC_bookkeeping}.}
\label{alg:kmc_ppmd}
\begin{algorithmic}[1]
    \STATEx{\textit{Set types \& proposed moves \texttt{(ParticleLoop)}}}
    \FORALL{charges $\partidxi=1,\dots,N$}
        \STATEx{\textit{\quad Set particle types based on position}}
        \STATE{$\mathcal{T}^{(\partidxi)} \mapsfrom T(\vec{r}^{(\partidxi)})$}
        \STATEx{\textit{\quad Set proposed positions and initialise masks}}
        \FOR{$k = 1, \dotsc, c_{\mathcal{T}^{(\partidxi)}}$}
            \STATE{$\mathcal{R}^{(\partidxi)}_k \mapsfrom \vec{r}^{(\partidxi)} + h\vec{\delta}^{\mathcal{T}^{(\partidxi)}}_{k}$}
            \STATE{$\mathcal{M}_k^{(\partidxi)} = 1$}
        \ENDFOR
    \ENDFOR
    \STATEx{\textit{Detect overlaps \texttt{(PairLoop)}}}
    \FORALL{pairs $(\partidxi,\partidxj)$ s.t.\ $|\vec{r}^{(\partidxi)} - \vec{r}^{(\partidxj)}| == h$}
        \FOR{$k = 1, \dotsc, c_{\mathcal{T}^{(\partidxi)}}$}
            \IF{$|{\vec{r}'}^{(\partidxi)}_k - \vec{r}^{(\partidxj)}| == 0$}
                \STATEx{\quad\quad\quad \textit{Flag proposed position as conflict}}
                \STATE{$\mathcal{M}_k^{(\partidxi)} = 0$}
            \ENDIF
        \ENDFOR
    \ENDFOR
\end{algorithmic}
\end{algorithm}
\section{Parallel Performance of standard FMM}\label{sec:results_parallelFMM}
To complement the results in \cite{Saunders2017a} and since our standard FMM implementation in itself might be of interest to others, we compare its performance and parallel scalability with the FFT accelerated Smooth Particle Mesh Ewald (SPME) approach in DL\_POLY\_4. Here we use a configuration which is based on the two ion NaCl ``TEST01'' \cite{Test01} scenario from the DL\_POLY test suite. The system is stabilised by adding a repulsive short range Lennard-Jones potential with a cutoff of $4\text{\AA}$. Due to this small cutoff the additional cost of the Lennard-Jones force calculation can be neglected in the reported runtimes. The initial configuration is a cubic lattice of alternating particle species with a lattice spacing of $3.3\text{\AA}$. To allow a fair comparison, for both implementations the parameters were adjusted such that both methods give comparable relative errors of $\sim10^{-6}$ on the total energy of the system; this required 10 expansion terms in the FMM implementation. For more details on the setup and quantification of errors see \cite{Saunders2018a}.

As for the results in Section \ref{sec:results_MPI}, all runs were carried out on the Intel Ivy Bridge E5-2650v2 nodes of the ``Balena'' HPC facility. In contrast to the setup in Tab. \ref{tab:node_layout}, the FMM code was run in two modes:
\begin{itemize}
  \item A pure MPI implementation, using distributed memory parallelism between the 16 cores of each node.
  \item A hybrid MPI+OpenMP mode with one MPI rank and 8 OpenMP threads per socket ($2\times 8$ OpenMP threads per 16-core node).
\end{itemize}
DL\_POLY was always run in pure MPI mode.
\subsection{Strong scaling}
To test the strong scalability we perform 200 Velocity Verlet integration steps of two systems containing $N=10^6$ and $N=4.0\cdot 10^6$ charged particles respectively.
\begin{figure}
\begin{center}
\begin{minipage}{0.45\linewidth}
\begin{center}
  \includegraphics[width=1.00\linewidth]{\figdir/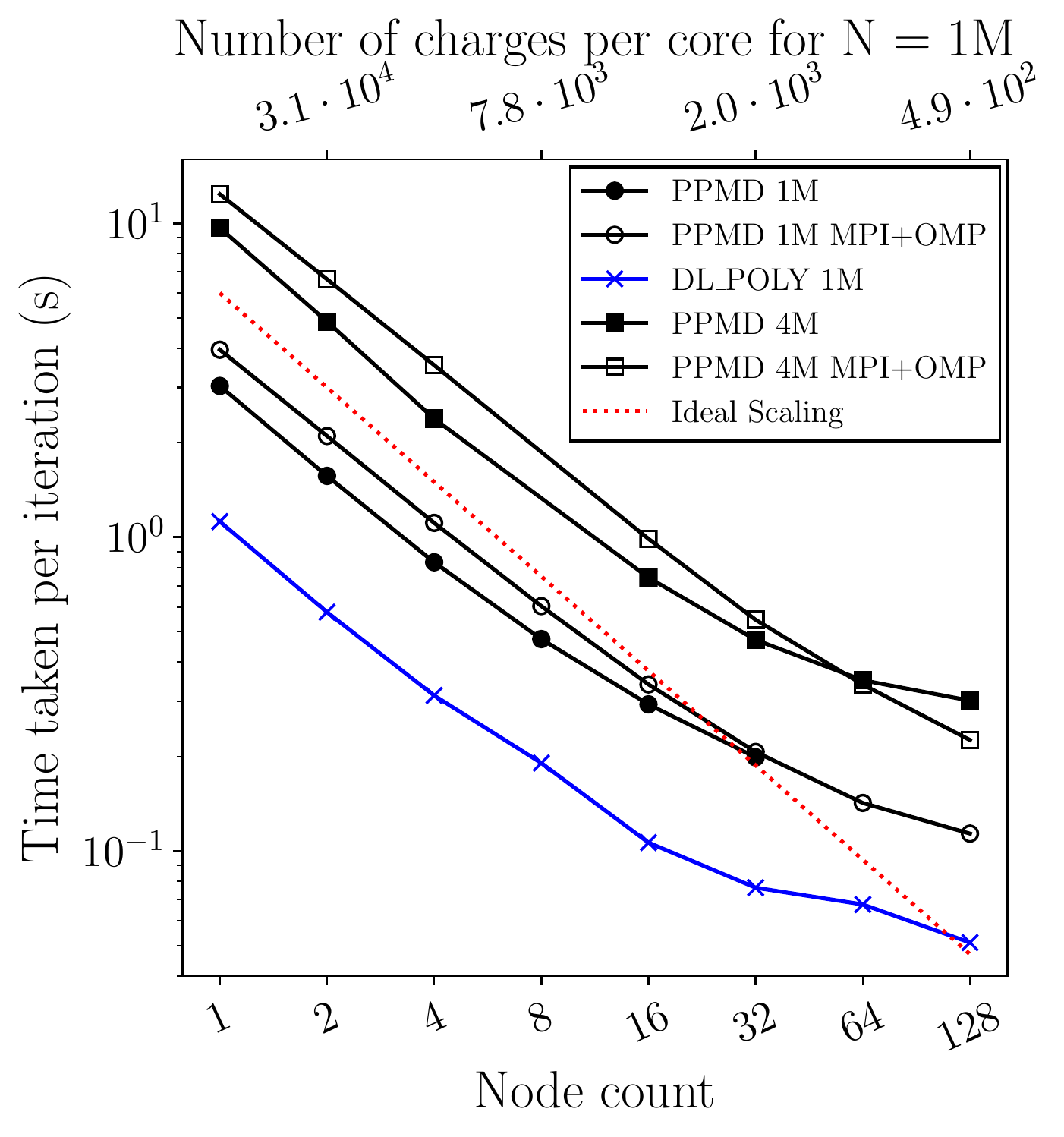}
\end{center}
\end{minipage}
\hfill
\begin{minipage}{0.45\linewidth}
\begin{center}
  \includegraphics[width=1.00\linewidth]{\figdir/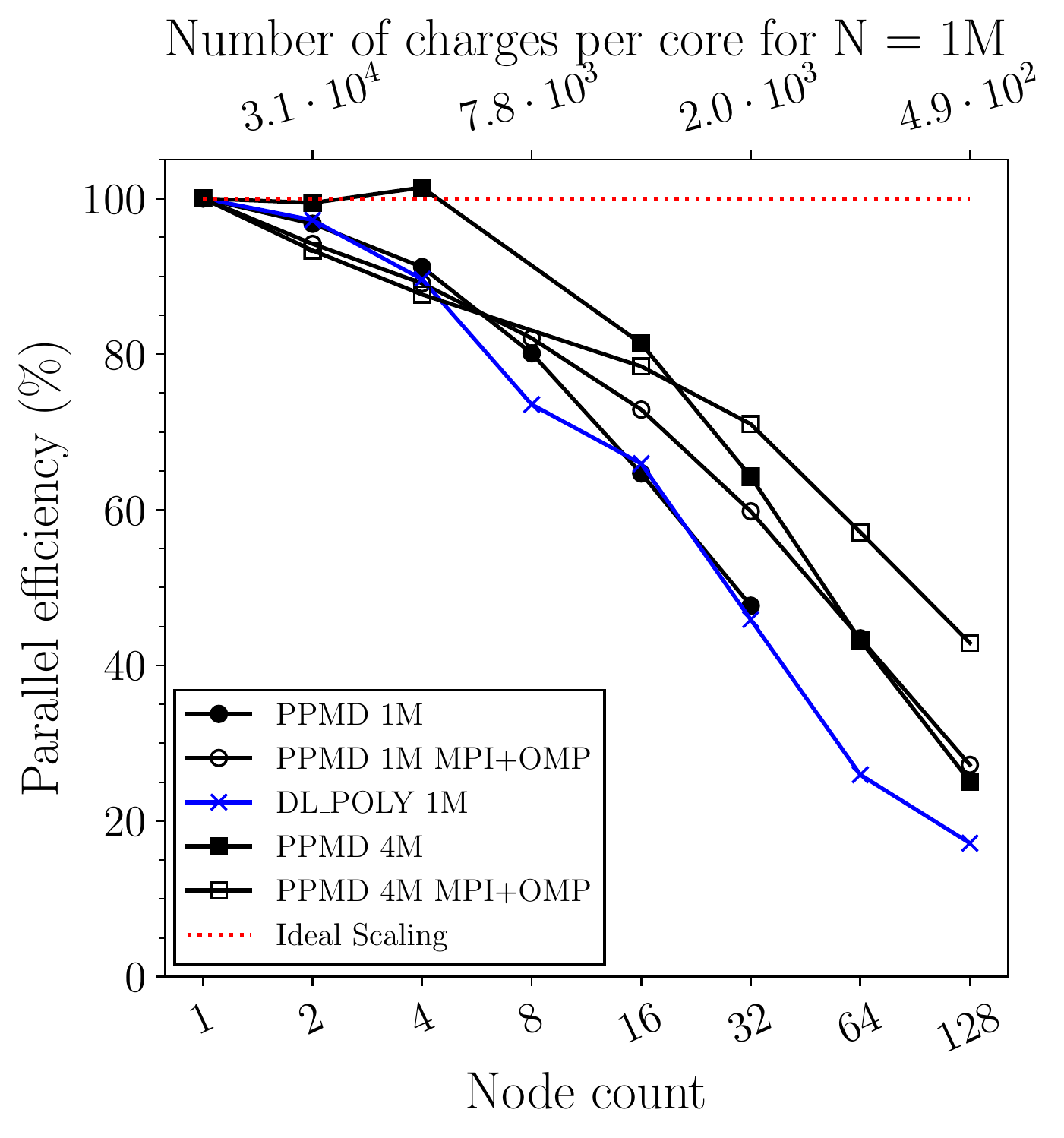}
\end{center}
\end{minipage}
\caption{Strong scaling comparison between our FMM implementation (labelled as ``PPMD'') and DL\_POLY FFT based SPME method. The time per Velocity Verlet step is shown for systems containing $N=10^6$ (1M) and $N=4.0 \cdot 10^6$ (4M) charges on the left. The strong parallel efficiency $E_S(P;N)$ as defined in Eq. \eqref{eqn:strong_eff}) is shown on the right.}
\label{fig:fmm_strong_scaling}
\end{center}
\end{figure}
The absolute runtimes in Fig. \ref{fig:fmm_strong_scaling} (left) demonstrate that the performance of our FMM implementation is in the same ballpark as the SPME algorithm in the mature DL\_POLY code, which is approximately $3\times$ faster. For larger core counts our FMM implementation does not exhibit unreasonable performance degradation; in fact it scales slightly better that the DL\_POLY code. This is further quantified by plotting the strong parallel scalability calculated according to Eq. \eqref{eqn:strong_eff} in Fig. \ref{fig:fmm_strong_scaling} (right). Running in MPI+OpenMP mode further increases parallel efficiency in the strong scaling limit. For smaller node counts, the efficiency of the hybrid approach is poorer than a for pure MPI setup. This is because OpenMP parallelisation introduces atomic operations not found in the distributed memory implementation. Those operations lead to reduced intra-node parallel efficiency, consistent with Amdahl's Law \cite{Amdahl1967}.
\subsection{Weak Scaling}
\begin{figure}
\begin{center}
\begin{minipage}{0.45\linewidth}
\begin{center}
  \includegraphics[width=1.00\linewidth]{\figdir/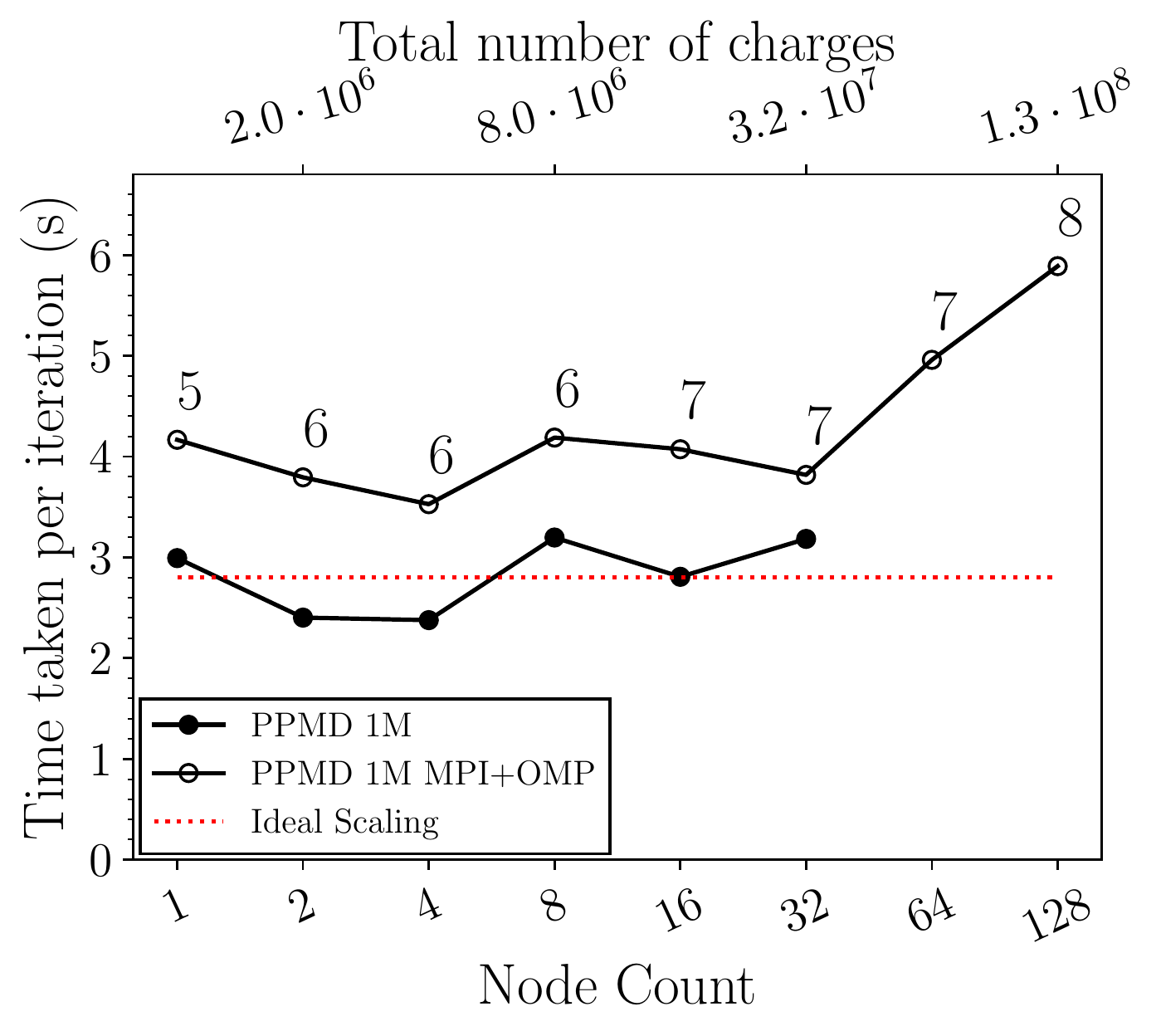}
\end{center}
\end{minipage}
\hfill
\begin{minipage}{0.45\linewidth}
\begin{center}
  \includegraphics[width=1.00\linewidth]{\figdir/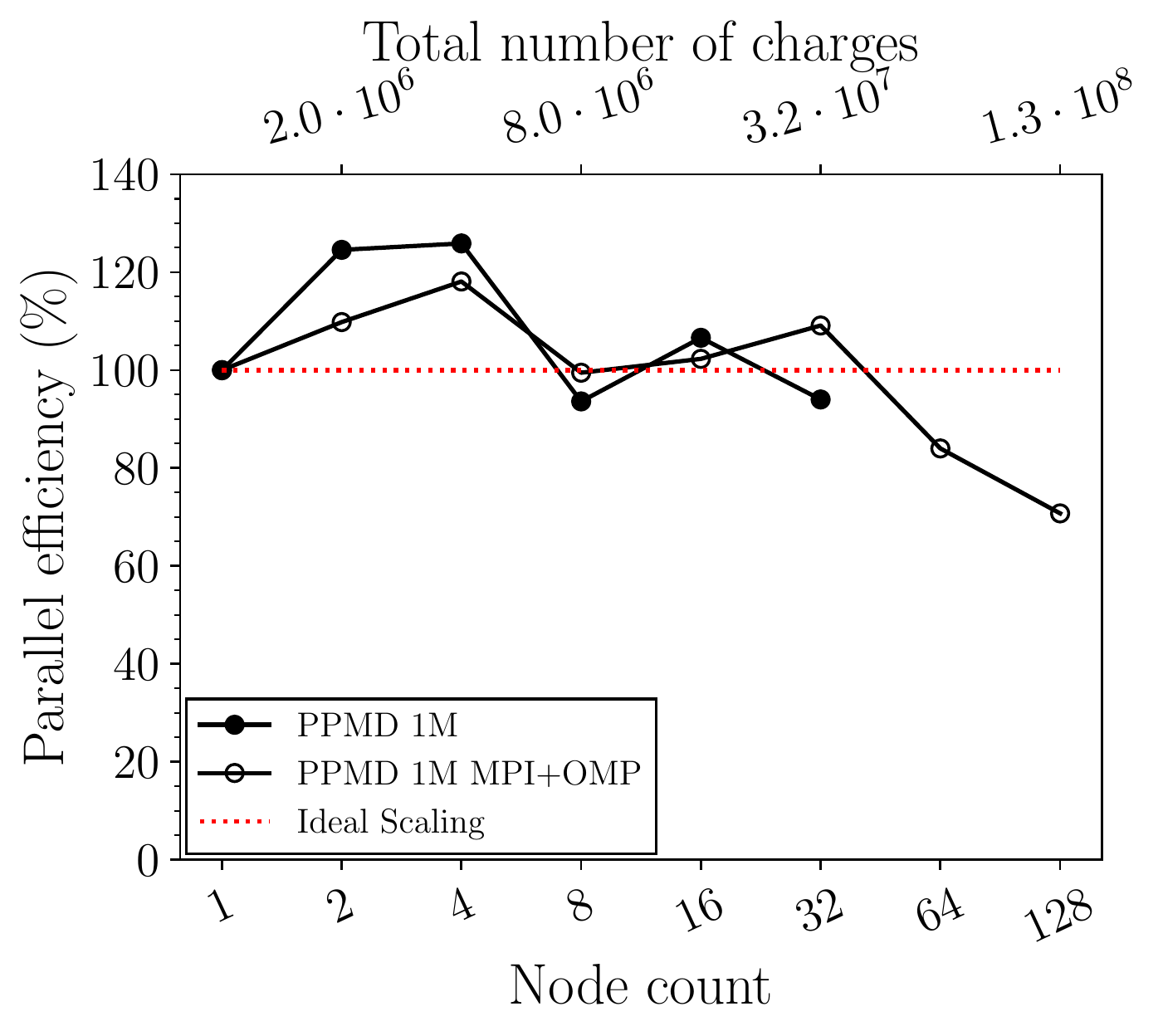}
\end{center}
\end{minipage}
\caption{Weak scaling of the time per Velocity Verlet step for the standard FMM implementation. The absolute time as a function of the number of nodes is shown on the left where floating numbers indicate the number of levels in the octal tree. The parallel efficiency as defined in Equation (\ref{eqn:weak_eff}) is shown on the right.}
\label{fig:fmm_weak_scaling}
\end{center}
\end{figure}

In the corresponding weak scaling experiment the number $N_{\text{local}}=10^6$ of charges per node is kept fixed, while the total number of charges $N=P\cdot N_{\text{local}}$ increases in proportion to the number of nodes $P$. Since the computational complexity of the FMM algorithm is proportional to $N$, we expect the time per FMM evaluation to be independent of $P$.
Fig. \ref{fig:fmm_weak_scaling} (left) shows the time per Velocity Verlet step for total problem sizes between $N=10^6$ and $N=1.28\cdot 10^8$. Due to memory inefficiencies in non-FMM related portions of code it was not possible to run the pure MPI implementation of the code on more than 32 nodes, and we only report results for the hybrid MPI+OpenMPI setup in this case. For each problem size the number of levels $L$ is adjusted to achieve optimal performance.
The parallel efficiency $E(P;N_{\text{local}})$ defined in Eq. \eqref{eqn:weak_eff} is plotted in Fig. \ref{fig:fmm_weak_scaling} (right). As expected, the time per Velocity Verlet step grows only slowly as the number of processors increases.

We refer the interested reader to \cite{Saunders2018a} for a further discussion of those results.
\FloatBarrier
\ifbool{PREPRINT}{ 
}{} 
\bibliographystyle{elsarticle-num}

\end{document}